\documentclass[aps,
               pra,
               twocolumn,
               showpacs
              ]{revtex4}

\usepackage{amsmath, amssymb}

\usepackage{pstricks}
\usepackage{graphicx}

\newcommand{\df}[1]{\mathrm{d}^{d+1}\! #1 \,}

\newcommand{\ddf}[1]{\mathrm{d}^{d}\! #1 \,}

\renewcommand{\vec}{\mathbf}

\newcommand{\eq}[1]{(\ref{eq:#1})}
\newcommand{\Eq}[1]{Eq.~(\ref{eq:#1})}

\newcommand{\Fig}[1]{Fig.~\ref{fig:#1}}

\newcommand{\Sect}[1]{Sect.~\ref{sec:#1}}

\newcommand{\Appendix}[1]{Appendix \ref{app:#1}}
\newcommand{\App}[1]{App.~\ref{app:#1}}

\newcommand{\OoN}{$1/\mathcal{N}$\ }

\begin{document}

\title{Matter Wave Turbulence: Beyond Kinetic Scaling}

\author{Christian~Scheppach$^1$}
\author{J{\"u}rgen~Berges$^{2,3}$}
\author{Thomas~Gasenzer$^{1,3}$}
\email{t.gasenzer@uni-heidelberg.de}
\affiliation{$^1$Institut f\"ur Theoretische Physik,
             Ruprecht-Karls-Universit\"at Heidelberg,
             Philosophenweg~16,
             69120~Heidelberg, Germany}
\affiliation{$^2$Institut f{\"u}r Kernphysik,
             Technische Universit{\"a}t Darmstadt,
             Schlo{\ss}gartenstr.~9,
             64289 Darmstadt, Germany}
\affiliation{$^{3}$ExtreMe Matter Institute (EMMI),
             GSI Helmholtzzentrum f\"ur Schwerionenforschung GmbH, 
             Planckstra\ss e~1, 
             64291~Darmstadt, Germany}


\begin{abstract}
Turbulent scaling phenomena are studied in an ultracold Bose gas away from thermal equilibrium. 
Fixed points of the dynamical evolution are characterized in terms of universal scaling exponents of correlation functions.
The scaling behavior is determined analytically in the framework of quantum field theory, using a nonperturbative approximation of the two-particle irreducible effective action. 
While perturbative Kolmogorov scaling is recovered at higher energies, scaling solutions with anomalously large exponents arise in the infrared regime of the turbulence spectrum.
The extraordinary enhancement in the momentum dependence of long-range correlations could be experimentally accessible in dilute ultracold atomic gases. 
Such experiments have the potential to provide insight into dynamical phenomena directly relevant also in other present-day focus areas like heavy-ion collisions and early-universe cosmology.
\end{abstract}


\maketitle

\section{Introduction}
\label{sec:Intro}
Turbulence is  a characteristic phenomenon naturally observed in many dynamical settings far from equilibrium. Besides its generic appearance in the context of fluid dynamics, wave turbulence has become an important topic in non-linear dynamics. In laboratory experiments of ultracold atoms, recently turbulent dynamics of vortical motion in Bose-Einstein condensates has received increasing attention from both theory and experiment, see Refs.~\cite{Berloff2002a,Parker2005a,Kobayashi2007a,Henn2009a}. Similar experiments may provide insight into dynamical phenomena relevant also in other present-day focus areas like collision experiments of heavy nuclei and reheating after inflation in early-universe cosmology, see Refs.~\cite{Strickland:2007fm,Micha:2004bv,Berges:2008pc} for recent reviews.

The theoretical framework for turbulence was formed, to a large part, in the 1940s~\cite{Kolmogorov1941a}. In the usual approach, wave turbulence is characterized by scaling laws derived in the framework of kinetic theory, see Refs.~\cite{Zakharov1992a,Micha:2004bv} for reviews with further references. There are two requirements for the validity of kinetic theory: The de Broglie wavelengths of particles must be small compared to the mean free path between collisions. This allows one to describe them as classical colliding particles. The second requirement is that the duration of individual scattering events should be small compared to the mean free time between collisions. Otherwise, interference between successive collisions would spoil their mutual independence. 

We study, in the present work, turbulent dynamics of an ultracold Bose gas beyond quantum kinetic theory. We use quantum field theoretical methods, with the aim of forming a theory framework for the systematic study of turbulence phenomena in experiments with ultracold atoms. 
Non-thermal scaling solutions for the dynamics are found, and these are characterized, in the infrared, by anomalously large exponents, confirming results recently presented in Refs.~\cite{Berges:2008wm,Berges:2008sr} in the context of relativistic field theory. 
Such an extraordinary enhancement in the low-energy distribution of matter-wave modes could be experimentally accessible in dilute ultracold atomic gases.

We point out in this work the strong connections between the phenomenon of "weak wave turbulence", which is perturbative and well described within kinetic theory, and the predicted non-perturbative low-energy scaling solutions. Since the latter occur in the deep infrared, the scaling behavior comes with distinctive properties such as critical slowing down and universality characteristic for critical phenomena. We explain in detail why this far-from-equilibrium critical dynamics may be understood as a "strong turbulence" phenomenon.  
As in the kinetic theory of weak wave turbulence, two different solutions are found in both the ultraviolet and infrared limits. 
It is shown that the obtained solutions corresponding to a quasiparticle cascade in the ultraviolet regime require a fixed dispersion relation between frequency and momentum while it is sufficient to have energy conservation for the solutions which relate to a scale-invariant energy flux, i.e., energy cascade.

Turbulence can arise as a stationary phenomenon in a driven system away from thermal equilibrium, as well as during a transient period after the driving force has ceased.
It then offers the system a special path of transit to eventual thermalization. Progress in understanding the structure of state space with respect to trajectories connecting non-equilibrium with equilibrium has potential applications in very different areas of physics.
The preparation of ultracold atomic Bose and Fermi gases in various trapping environments allows to precisely study quantum many-body dynamics of strongly correlated systems.
A number of experiments have recently focused on far-from-equilibrium dynamics, long-time evolution, and thermalization in such systems, cf., e.g., Refs.~\cite{Kinoshita2006a,Sadler2006a,Hofferberth2007a}.    
    
A possible realization of turbulent dynamics in an ultracold atomic gas could be achieved with a Bose-Einstein condensate mode as a source of excitations, which starts oscillating parametrically after a quench of the scattering length and thus the chemical potential.
The associated phenomenon of parametric resonance leads to a strong enhancement of infrared modes. This has been studied in great detail in weakly coupled theories of preheating after inflation in the early universe~\cite{Traschen:1990sw,Kofman:1994rk,Khlebnikov:1996mc,Berges:2002cz,Micha:2002ey,Berges:2008wm}. In this case, weak turbulence describes above a characteristic momentum scale the gradual transfer of energy towards higher momenta in the form of a cascade. 

In contrast, for low momenta occupation numbers grow non-perturbatively large and exhibit a different scaling behavior. A correct description of these low-energy modes requires non-perturbative approximations. A non-perturbative approach that has gained substantial success in describing far-from-equilibrium evolution over long times is based on the 2PI effective action \cite{Luttinger1960a,Baym1962a,Cornwall1974a} in next-to-leading order (NLO)$1/\mathcal{N}$ approximation \cite{Berges:2001fi,Aarts:2002dj}.
Dynamic equations derived from an nPI effective action by definition obey crucial conservation laws, as for the total energy and, in a non-relativistic system, for particle number.
Here we use the 2PI approach in NLO \OoN approximation to study turbulence in ultracold gases.

This article is organised as follows: 
\Sect{2PIEA-Approach} provides the set of non-equilibrium dynamic equations which are used to find a necessary condition for the existence of a non-thermal stationary point.
In \Sect{NonthFP}, we derive the stationarity condition and discuss the thermal fixed point and summarize results for turbulent scaling found within kinetic theory as far as relevant for the later discussion.
\Sect{TurbDynScaling} provides the derivation of the scaling behavior, both in the perturbative regime as well in the non-perturbative infrared limit of low-energy mode excitations.   
Our conclusions are drawn in \Sect{Discussion}.
Technical details and a short summary of the real-time functional quantum field theory are provided in the appendix.

\section{Nonequilibrium dynamics beyond kinetic theory}
\label{sec:2PIEA-Approach}
In this section we summarize the aspects of nonequilibrium quantum field theory most relevant to the discussion of turbulence in this article.
Our analysis is based on the two-particle irreducible (2PI) effective action \cite{Luttinger1960a,Baym1962a,Cornwall1974a} applied to real-time dynamics.
We approximate the 2PI effective action to next-to-leading (NLO) in the nonperturbative expansion in powers of the inverse number $\mathcal{N}$ of internal degrees of freedom \cite{Berges:2001fi,Aarts:2002dj}. This description can reach substantially beyond a quantum kinetic approach, which is typically restricted to on-energy-shell scattering between quasiparticles with a well-defined dispersion relation.
After successful applications of these nonperturbative expansions to the study of far-from-equilibrium dynamics and thermalization in relativistic bosonic~\cite{Berges:2001fi,Berges:2002cz, Cooper:2002qd, Arrizabalaga:2004iw} and fermionic~\cite{Berges:2002wr, Berges:2004ce, Berges:2009bx} theories, they have recently been employed in the context of ultracold bosonic quantum gases~\cite{Rey2004a,Gasenzer:2005ze,Temme2006a,Berges:2007ym, Branschadel:2008sk}.
For introductory texts see, e.g., Refs.~\cite{Berges:2004yj, Gasenzer2009a}.
Some details about the 2PI effective action approach to nonequilibrium dynamics relevant for our discussion can also be found in App.~\ref{app:2PIEA-Approach}.

We consider the evolution of an ultracold bosonic quantum many body system described by the complex $\cal N$-component Heisenberg field operators $\Phi_{\alpha}(t,\mathbf{x})$, $\alpha=1,\ldots,\mathcal{N}$ in $d$ spatial dimensions, obeying the commutation relations $[\Phi_{\alpha}(t,\mathbf{x}),\Phi^\dagger_{\beta}(t,\mathbf{y})]=\delta_{\alpha\beta}\delta(\mathbf{x}-\mathbf{y})$, $[\Phi_{\alpha}(t,\mathbf{x}),\Phi_{\beta}(t,\mathbf{y})]=0$.
In the following we choose a basis where the field is written in terms of its real and imaginary components, $\Phi_{\alpha}=(\Phi_{1,\alpha}+i\Phi_{2,\alpha})/\sqrt{2}$.
Including the field component index as well as the ``magnetic'' index $\alpha$ into a single index $a=(i_{a},\alpha)$, with $i_{a}=1,2$, the commutation relations read 
\begin{align}
  [{\Phi}_a(t,{\bf x}),{\Phi}_b(t,{\bf y})]=-\sigma^2_{i_{a}i_{b}}\delta_{\alpha\beta}
  \delta({\bf x} -{\bf y}),
  \label{eq:BoseCommutator}
\end{align}
where $\sigma^{2}$ denotes the Pauli $2$-matrix \footnote{We use natural units where $\hbar=1$.}.
We consider a quantum field theory for a complex $\cal N$-component field $\varphi_a(x)$  ($a=(i_{a},\alpha)$, $i_{a}=1,2$, $\alpha=1,...,{\cal N}$) with quartic interactions,
\begin{align}
  S[\varphi]
  &= \frac{1}{2}\int_{x y} \varphi_a(x) iD_{ab}^{-1}(x,y)\varphi_b(y)
  \nonumber\\
  & -\frac{g}{4{\cal N}} \int_{x} \varphi_a(x)\varphi_a(x)\varphi_b(x)\varphi_b(x) ,
\label{eq:Sclassphi4}
\end{align}
where we use the notation $\int_x \equiv \int \mathrm{d} x_0 \int \mathrm{d}^d x$ with $(x_0,\mathrm{x}) = (t,\mathrm{x})$. 
This model describes, e.g., an ultracold Bose gas of atoms with $\mathcal{N}$ hyperfine sublevels whose interaction strength $g$ does not depend on the particular hyperfine scattering channel of a pair of atoms.
In $d=3$ dimensions the coupling strength in such a system is $g=4\pi a/m$,  $a$ being the $s$-wave scattering length.
The free inverse classical propagator of this model reads
\begin{align}
\label{eq:G0inv}
  &iD^{-1}_{ab}(x,y)
  =\left.iG^{-1}_{0,ab}(x,y)\right|_{\phi=0}
  \nonumber\\
  &\quad=
   \delta(x-y)\delta_{\alpha\beta}\left[-i\sigma^2_{i_{a}i_{b}}\partial_{x_0}
   -H_\mathrm{1B}(x)\delta_{ab}
     \right],
\end{align}
see \Eq{G0}.
Here $H_\mathrm{1B}(x)=-\sum_{j=1}^{d}\partial^2_j/2m+V(x)$ denotes the
single-particle Hamiltonian, and we choose, in the following, the external potential $V(x)$ to vanish.

\subsection{Dynamic equations}
\label{sec:DynEqs}
In this section we recall the equations which describe the time evolution of the lowest order connected correlation functions or cumulants \cite{Berges:2007ym,Gasenzer2009a}
\begin{align}
\label{eq:phi}
  \phi_{a}(x) 
  &= \langle\Phi_{a}(x)\rangle,
  \\
  \label{eq:G}
  G_{ab}(x,y)
  &= \langle\mathcal{T}\Phi_{a}(x)\Phi_{b}(y)\rangle - \phi_{a}(x)\phi_{b}(y).
\end{align}
The two-time correlation function $G$ defined in \Eq{G} involves a time-ordered product of in general noncommuting field operators. Therefore, $G$ can be decomposed as~\cite{Aarts:2001qa}
\begin{equation}
\label{eq:GitoFrho}
  G_{ab}(x,y) = F_{ab}(x,y) -\frac{i}{2}\mathrm{sgn}(x_0-y_0)\rho_{ab}(x,y),
\end{equation}
where the signum function $\mathrm{sgn}(x_0-y_0)$ evaluates to $1$ ($-1$) for $x_{0}$ later (earlier) than $y_{0}$, and where the statistical component $F$ and the spectral part $\rho$ are defined in terms of the anticommutator and commutator of the fields, respectively,
\begin{align}
\label{eq:Fandrho}
  F_{ab}(x,y)
  &=\mbox{$\frac{1}{2}$}\langle\{\Phi_a(x),\Phi_b(y)\}\rangle_c,
  \\
  \rho_{ab}(x,y)
 & =i\langle[\Phi_a(x),\Phi_b(y)]\rangle.
\end{align}
Here, $\langle\cdot\rangle_{c}$ is a short-hand notation for the cumulant $\langle \Phi_{a}\Phi_{b}\rangle_{c}=\langle \Phi_{a}\Phi_{b}\rangle-\langle \Phi_{a}\rangle\langle\Phi_{b}\rangle$, see \Eq{G}.

The resulting integro-differential dynamic equations for $\phi$, $F$, and $\rho$ read for Gaussian initial conditions \cite{Berges:2007ym,Gasenzer2009a}
\begin{widetext}
\begin{align}
 &   \Big(-i\sigma^2_{ab}\partial_{x_0}
    -g\,F_{ab}(x,x)\Big)\phi_b(x) -\Big(H_\mathrm{1B}(x)
    +\frac{g}{2}\big[\phi_c(x)\phi_c(x)
    +F_{cc}(x,x)\big]\Big)
    \phi_a(x) 
   = \int_{t_0}^{x_0} \!\mathrm{d}y\,
   \Sigma^\rho_{ab}(x,y;\phi\equiv 0)\,\phi_b(y) ,
\label{eq:EOMphi}
  \\
  &\left[i\sigma^2_{ac}\partial_{x_0} + M_{ac}(x) \right]
    F_{cb}(x,y)
  = - \int_{t_0}^{x_0} \! \mathrm{d} z\,
    \Sigma^{\rho}_{ac}(x,z;\phi) F_{cb}(z,y)
  + \int_{t_0}^{y_0} \! \mathrm{d}z\,
    \Sigma^{F}_{ac}(x,z;\phi) \rho_{cb}(z,y) ,
\label{eq:EOMF}
  \\[1.5ex]
  &
  \left[i\sigma^2_{ac}\partial_{x_0}
    +M_{ac}(x) \right] \rho_{cb} (x,y)  
  = -\int_{y_0}^{x_0} \! \mathrm{d}z\,
  \Sigma^{\rho}_{ac} (x,z;\phi)
  \rho_{cb}(z,y).
\label{eq:EOMrho}
\end{align}
These identities are equivalent to Kadanoff-Baym or Schwinger-Dyson equations.
Here we employ the notation $\int_{t}^{t'}\mathrm{d}z = \int_t^{t'}\mathrm{d}z_0\int \mathrm{d}^dz$.
The ``mass'' matrix $M$ containing the free Hamiltonian as well as mean-field potential terms is defined as
\begin{equation}
  M_{ab}(x)
   = \delta_{ab}
  \Big[H_\mathrm{1B}(x)
   + \frac{g}{2}\Big(\phi_c(x)\phi_c(x)+F_{cc}(x,x)\Big)\Big]
   + g\Big(\phi_a(x)\phi_b(x)+F_{ab}(x,x)\Big).
\label{eq:MM}
\end{equation}
\end{widetext}
The self energy $\Sigma$ acts as the kernel in the non-Markovian memory integrals in the above dynamic equations and accounts for collisions building up correlations in the system. It is obtained as the derivative of the 2PI part $\Gamma_{2}$,
\begin{equation}
\label{eq:SigmafromGamma2}
  \Sigma_{ab}(x,y;\phi,G)=2i\frac{\delta\Gamma_2[\phi,G]}{\delta G_{ab}(x,y)},
\end{equation}
and has been decomposed into a local mean-field part $\Sigma^{(0)}_{ab}(x)$ adding to the mass matrix, and a nonlocal part written in terms of statistical and spectral components,
\begin{align}
\label{eq:Sigma0Frho}
  \Sigma_{ab}(x,y)
  &=\Sigma^{(0)}_{ab}(x)\delta (x-y)
  \nonumber\\
  &+\ \Sigma^F_{ab}(x,y)-\frac{i}{2}\mathrm{sgn}(x_0-y_0)\Sigma^\rho_{ab}(x,y).
\end{align}
These non-local parts form the kernels for the memory integrals on the right-hand sides of the integro-differential dynamic equations \eq{EOMphi}-\eq{EOMrho}. 

\subsection{NLO 2PI $1/{\cal N}$ expansion}
\label{sec:NLO1N}
To practically solve the dynamic equations \eq{EOMphi}-\eq{EOMrho}, details about the self energy $\Sigma$ are required, and these are, in general, only available to a certain approximation. In the following we employ
an expansion of $\Gamma_2$ in powers of the inverse number of field degrees of freedom $\cal N$ \cite{Berges:2001fi,Aarts:2002dj,Berges:2004yj}.
The $1/{\cal N}$  expansion to next-to-leading order (NLO) is equivalent to replacing certain vertices in a loop expansion by a bubble-resummed vertex \cite{Aarts:2002dj,Gasenzer:2005ze}.
In the context of an ultracold Bose gas, it has been discussed in Refs.~\cite{Gasenzer:2005ze,Temme2006a,Berges:2007ym,Branschadel:2008sk}.
This approximation scheme has also been recovered in a functional renormalization group inspired approach \cite{Gasenzer:2008zz} where it results as a truncation in orders of proper $n$-point functions combined with an $s$-channel approximation of the equation for the proper four-vertex.

\begin{figure}[tb]
\begin{center}
\resizebox{0.9\columnwidth}{!}{
\includegraphics{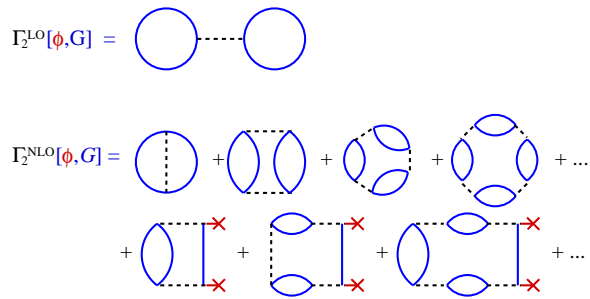}
}
\end{center}
\vspace*{-3ex}
\caption{
(Color online) Diagrammatic representation of the  leading order (LO) and next-to-leading order (NLO) contributions in the $1/\cal N$-expansion, to the 2PI part $\Gamma_2[\phi,G]$ of the 2PI effective action.
The thick blue lines represent 2-point functions $G_{ab}(x,y)$, the red crosses field insertions $\phi_a(x)$, and the dashed lines vertices $g/4\mathcal{N}\delta(x-y)$.
At each vertex, it is summed over double field indices $a$ and integrated over double time and space variables $x$.
}
\label{fig:DiagrExpGamma2NLO}
\end{figure}

In this scheme, the contribution $\Gamma_2[\phi,G]$ to the 2PI effective action involves a leading (LO) and next-to-leading order (NLO) part which can be diagrammatically represented as shown in \Fig{DiagrExpGamma2NLO} in terms of 2PI closed loop diagrams involving only bare vertices, full propagators $G$, and field insertions $\phi$.
While the leading-order contribution involves one diagram, in NLO a chain of bubble diagrams is resummed.
All of these diagrams are proportional to the same power of $1/\cal N$ since each vertex scales with $1/\cal N$, which is cancelled by the (blue) propagator loops which scale with $\cal N$ since they involve a summation over the field indices from $1$ to $\cal N$ \cite{Gasenzer:2005ze,Gasenzer2009a}.

Inserting the 2PI effective action expanded to NLO in $1/\mathcal{N}$ into \Eq{SigmafromGamma2} one calculates the self energy $\Sigma$ and from this the local and nonlocal contributions defined in \Eq{Sigma0Frho}.
The resulting expressions are provided in \Appendix{Sigma}.

\section{Stationary solutions of the evolution equations}
\label{sec:NonthFP}
In this section, we set up the condition for stationary solutions of the dynamic equations introduced above which is later used to derive turbulent scaling behavior. 
We furthermore discuss the thermal equilibrium solutions as well as the scaling solutions associated to weak wave turbulence found in the framework of kinetic theory.

\subsection{Stationarity condition}
We consider stationary homogeneous solutions of the dynamic equations \eq{EOMphi}-\eq{EOMrho}, i.e., solutions invariant under translations in time and space,
\begin{align}
  \phi_a(x)
  =\phi_{a}(t) 
  &\sim \exp(-i\mu t)\,;
  \nonumber\\ 
  F_{ab}(x,y)=F_{ab}(x-y);
  \quad &
  \rho_{ab}(x,y)=\rho_{ab}(x-y)
\label{eq:invariant}
\end{align}
Note that the stationarity condition for $\phi$ allows for a remaining rotating phase with constant angular velocity corresponding to a chemical potential, i.e., it only requires the condensate density $|\phi|^{2}$ to be constant in time.

As we are looking for stationary solutions, we can send the initial time in the dynamic equations to $t_0 \rightarrow -\infty$.
Consequently, also the self-energy components $\Sigma^{F,\rho}_{ab}$, as well as the functions $I^{F,\rho}$ and $P^{F,\rho}$ which are defined in Eqs.~\eq{SigmaNLO1N}-\eq{HFHrho} depend only on the relative coordinate, $\Sigma^\rho_{ab}(x,y)=\Sigma^\rho_{ab}(x-y)$, etc. 
We write the Fourier transform of the statistical correlation function as $F_{ab}(p) := \int\df{x}\exp(ipx)F_{ab}(x)$, etc.~for all other two-point functions.

A necessary condition for a solution of the dynamic equation \eq{EOMF} for the statistical correlation function $F$ to be translationally invariant as defined in \Eq{invariant} 
reads~\cite{Berges:2008wm}
\begin{equation}
  J(p) \equiv \Sigma^\rho_{ab}(p)F_{ba}(p) - \Sigma^F_{ab}(p)\rho_{ba}(p) \stackrel{!}{=} 0.
\label{eq:stationarity}
\end{equation}
In \Appendix{StatCond} we derive the necessity of condition \eq{stationarity} for stationarity of $F$, using the symmetry properties of the correlation functions involved.
This stationarity condition can also be obtained from the trace of a gradient expansion of the evolution equation (\ref{eq:EOMF}), which leads to $2\omega\partial_{X_{0}} F(X_{0},\omega,\vec p) = J(X_{0},\omega,\vec p)$, with $\omega \equiv p_{0}$~\cite{Berges:2005md}. In situations where $F$ and $\rho$ are independent of the time $X_{0}=(x_0+y_0)/2$, 
the gradient expansion in $X_{0}$ involved in the derivation of this equation becomes exact.
We remark that there is no analogous condition following from the dynamic equation \eq{EOMrho} for $\rho$. Following the same arguments as for $F$, we find that \Eq{EOMrho} allows for translationally invariant solutions obeying the conditions \eq{invariant}, see \Appendix{StatCond}.

\subsection{Thermal equilibrium}
\label{sec:ThFP}
Before proceeding to nonthermal stationary solutions we verify that in thermal equilibrium the condition (\ref{eq:stationarity}) is satisfied: 
In this limit, the grand canonical density matrix reads $\hat \rho = {\exp(-\beta(\hat H-\mu\hat N)}/\mathrm{tr}\exp[-\beta (\hat H-\mu\hat N)]$, and one can deduce the fluctuation-dissipation relation, 
\begin{equation}
  F^{(\mathsf{th})}_{ab}(\omega,\vec p) 
  = -i\left(n_{\mathrm{BE}}(\omega) + \frac{1}{2}\right)
  \rho^{(\mathsf{th})}_{cb}(\omega,\vec p)
\label{eq:f-d-rel}
\end{equation}
with
\begin{equation}
  n_\mathrm{BE}(\omega) = {1}/({e^{\beta(\omega-\mu)}-1}),
\label{eq:nBE}
\end{equation}
$\omega=p_{0}$.
For a detailed discussion see, e.g., \cite{Berges:2004yj,Gasenzer2009a}:
In addition to this relation there is a similar relation for the self-energies,
\begin{equation}
  \Sigma^{F(\mathsf{th})}_{ab}(\omega,\vec p) 
  = -i\left(n_{\mathrm{BE}}(\omega) + \frac{1}{2}\right)
  \Sigma^{\rho(\mathsf{th})}_{ab}(\omega,\vec p).
\end{equation}
Substituting this equation and \Eq{f-d-rel} into \Eq{stationarity} one finds, as expected, in thermal equilibrium the stationarity condition is fulfilled.

\subsection{Weak wave turbulence from kinetic equations}
\label{sec:KinFP}
In this subsection we briefly review the Kolmogorov theory of turbulent scaling solutions of the wave kinetic equation \cite{Zakharov1992a}. 
Going away from thermal equilibrium, a kinetic description of the time evolution implies a quasiparticle assumption, which assumes a fixed dispersion relation $\omega=\omega(\vec p)$ between momentum and frequency. Similar to (\ref{eq:f-d-rel}) one writes for spatially homogeneous situations, with $t \equiv X^0$,
\begin{equation}
  F^{(\mathsf{qp})}_{ab}(\omega,\vec p) 
  = -i\left(n^{(\mathsf{qp})}(\omega) + \frac{1}{2}\right)
  \rho^{(\mathsf{qp})}_{ab}(\omega,\vec p) ,
\label{eq:qp-ansatz}
\end{equation}
where the quantity $n^{(\mathsf{qp})}(\omega)$ will play the role of a quasiparticle number of mode $\vec p$ for fixed $\omega=\omega(\vec p)$.
The spectral function is given, e.g., for a cold Bose gas in the symmetric, i.e., non-condensed phase, with quadratic dispersion $\omega(\vec p)=\vec p^{2}/2m$, by the ideal-gas expression (coupling $g=0$) \cite{Branschadel:2008sk}
\begin{align}
  \rho_{ab}^{(\mathsf{qp})}(p) =\rho_{ab}^{(\mathsf{0})}(p)
  &= i\pi\left(\delta(\omega - \frac{{\vec p}^2}{2m}) + \delta(\omega + \frac{{\vec p}^2}{2m})\right)
  \nonumber\\
  &\quad\times\ [ \mathrm{sgn}(\omega)\,\delta_{ab}-\sigma^{2}_{ab}],
\label{eq:freerho}
\end{align}
where $\sigma^{2}$ is the Pauli 2-matrix. 

Inserting this ansatz into the dynamic equation \eq{EOMF} for $F$, where the self energy has been approximated by the second-order-in-$g$ expressions given in Appendix \ref{app:Sigma} one derives, in leading order of a gradient expansion in $t=X_{0}$, and for $\phi=0$, the quantum four-wave kinetic equation \cite{Berges:2004yj,Branschadel:2008sk}
\begin{align}
  \partial_{t}n_{\vec p}
  &= I(\vec p,t),
  \label{eq:QKinEq}
  \\
   I(\vec p,t)
  &= g^{2}\int\ddf{k}\ddf{q}\ddf{r}|T_{\vec p\vec k\vec q\vec r}|^{2}\delta(\vec p+\vec k -\vec q - \vec r)
  \nonumber\\
  &\qquad\quad\times\
  \delta(\omega_{\vec p}+\omega_{\vec k}-\omega_{\vec q}-\omega_{\vec r})
  \nonumber\\
  &\qquad\quad\times\
  [(n_{\vec p}+1)(n_{\vec k}+1)n_{\vec q}n_{\vec r}
  \nonumber\\
  &\qquad\qquad -\
  n_{\vec p}n_{\vec k}(n_{\vec q}+1)(n_{\vec r}+1)],
  \label{eq:KinScattInt}
\end{align}
where $n_{\vec p}\equiv n^{(\mathsf{qp})}(\vec p,t)$.
In the case defined by \Eq{freerho}, the transition matrix element squared $|T_{\vec p\vec k\vec q\vec r}|^{2}$ is a numerical constant independent of momenta. The perturbative expansion for weak coupling $g$ underlying (\ref{eq:KinScattInt}) restricts the occupation numbers to be parametrically $n_{\vec p} \ll (|\mathbf{p}|a)^{-1}$. 
For $1 \ll n_{\vec p} \ll (|\mathbf{p}|a)^{-1}$, where $a$ is the $s$-wave scattering length, the equation goes over to the classical kinetic equation. This is the regime, where scaling solutions describing Kolmogorov wave turbulence can be obtained. In the quantum limit $n_{\vec p}\ll1$, waves behave like particles, and \Eq{QKinEq} reduces to the Boltzmann equation. This is typically the range of high momenta, where occupation numbers are low, and no turbulent scaling will be observed in the quantum regime. Owing to the local conservation of $n_{\vec p}$, the above kinetic equation can be written as a continuity equation
\begin{align}
  \partial_{t}n^{(\mathsf{qp})}(\vec p,t) + \partial_{i}Q_{i}(\vec p,t) =0,
\end{align}
with the divergence of the current, $\partial_{i}Q_{i}(\vec p,t)=-I(\vec p,t)$, defined in terms of the scattering integral $I$, \Eq{KinScattInt}.

Alternatively, one can write the kinetic equation \eq{QKinEq} as a continuity equation for the energy density $\varepsilon(\vec p,t)=\omega(\vec p)n^{(\mathsf{qp})}(\vec p,t)$, with the divergence of the current $\mathbf{P}$ of energy flow given by the integral $\partial_{i}P_{i}(\vec p,t)=-\mathcal{I}(\vec p,t)=-\omega(\vec p)I(\vec p,t)$.

Taking into account the spherical symmetry of the interactions, the scattering integral $I(\vec p,t)$ can be averaged over the spatial directions which returns an integral $I(\omega({\vec p}))$ depending on the frequency only.
As is shown in detail in Ref.~\cite{Zakharov1992a}, the stationarity condition $I(\omega({\vec p}))=0$ has four universal scaling solutions 
($\omega(\vec p)\equiv\omega(|\vec p|)$) for the classical kinetic equation:
\begin{align}
  n^{(\mathsf{qp})}(\omega(s\vec p)) \equiv n^{(\mathsf{qp})}(s|\vec p|)=s^{-\kappa}n^{(\mathsf{qp})}(|\vec p|),
\end{align}
where the exponent is either $\kappa=0$ and $\kappa=z$, for the constant and thermal solutions, respectively, $z$ being the scaling exponent of the quasiparticle frequency, $\omega(s\vec p)=s^{z}\omega(\vec p)$, or
\begin{align}
  \kappa &= \kappa_{Q} = \frac{1}{3}(3d+2m-z),
  \\
  \kappa &= \kappa_{P} = \kappa_{Q} + \frac{z}{3}.
\end{align}
Here, $m$ is the scaling exponent of the transition matrix element, $T({s\vec p, s\vec k, s\vec q, s\vec r})=s^{m}T({\vec p,\vec k,\vec q,\vec r})\equiv s^{m}T_{\vec p\vec k\vec q\vec r}$, which evaluates to $m=0$ for the weakly interacting cold Bose gas away from unitarity.
The above nonthermal scaling exponents characterize the Kolmogorov stationary distributions of wave turbulence.
The solution with exponent $\kappa_{Q}$ corresponds to a $|\vec p|$-independent radial particle flux $|\vec p|^{d-1}Q(|\vec p|)$, while the solution with exponent $\kappa_{P}$ gives a momentum-independent energy flux $|\vec p|^{d-1}P(|\vec p|)$ \cite{Zakharov1992a}.
For the non-condensed cold Bose gas, one finds the momentum scaling exponents
\begin{align}
  \kappa_{Q} &= d-\frac{2}{3},
  \label{eq:KolQScaling}
  \\
  \kappa_{P} &= d.
  \label{eq:KolPScaling}
\end{align}
For a system with linear dispersion, $z=1$ and dominant four-wave interaction with $m=-2z=-2$, e.g., an ultracold Bose gas in the unitary limit at large momenta $|\vec p|$ (neglecting inelastic scattering), or a relativistic scalar theory in the high-energy limit \cite{Berges:2004yj}, one recovers the scaling exponents
\begin{align}
  \kappa_{Q} &= d-\frac{5}{3},
  \label{eq:KolQScalingzOne}
  \\
  \kappa_{P} &= d-\frac{4}{3},
  \label{eq:KolPScalingzOne}
\end{align}
which in $d=3$ evaluate to $\kappa_{Q} = {4}/{3}$ and $\kappa_{P} = {5}/{3}$ \footnote{%
We remark that the exponent $5/3$ is not to be confused with that in the well-known Kolmogorov-Obukhov ``$5/3$-scaling law'' \cite{Kolmogorov1941a,Obukhov1941a} which rather applies to the scaling of the radial energy spectrum $E(|\vec p|)\propto |\vec p|^{d-1}\omega(|\vec p|)n(|\vec p|)$  for turbulence phenomena like vorticity in an isotropic incompressible fluid where the density $\rho$ is the only relevant parameter.
In this case, the spectrum $E(|\vec p|)$ can be expressed in terms of $\rho$, the energy flux $P$, and the momentum $p$, in $d=3$ dimensions as $E(|\vec p|)\propto P^{2/3} \rho^{1/3} |\vec p|^{-5/3}$.
For wave turbulence which we discuss in this article, the frequency $\omega$ introduces, for each momentum $p$, a further relevant parameter.
The Kolmogorov-Obukhov law here results for $m=2$, which gives $n(|\vec p|)\sim |\vec p|^{-(d+10)/3}$ and, taking into account that, for dimensional reasons, the frequency follows the proportionality $\omega\propto(P|\vec p|^{5-d}/\rho)^{1/3}$, the spectral scaling $E(|\vec p|)\propto P^{2/3} \rho^{1/3}|\vec p|^{(d-8)/3}$, see, e.g., Ref.~\cite{Zakharov1992a}.}.

\subsection{Stationarity condition for the full dynamic equation}
\label{sec:StatCondNLO1N}
We proceed by considering the stationarity condition \eq{stationarity} beyond the above kinetic approximation, i.e., using the $1/{\cal N}$ expansion of the 2PI effective action to NLO following Refs.~\cite{Berges:2008wm,Berges:2008sr}. A major restriction of the kinetic description is that it cannot describe high occupation numbers $n_{\vec p}$ beyond the range of validity of perturbation theory. Since scaling solutions $n_{\vec p} \sim |{\vec p}|^{-\kappa}$ imply large occupancies at low momenta, kinetic theory breaks down in the infrared. In contrast, the non-perturbative 2PI $1/{\cal N}$ expansion to NLO allows us to investigate the low-momentum regime, which is characterized by different scaling solutions than the above Kolmogorov results.

\begin{figure}[tb]
\begin{center}
\resizebox{0.9\columnwidth}{!}{
\includegraphics{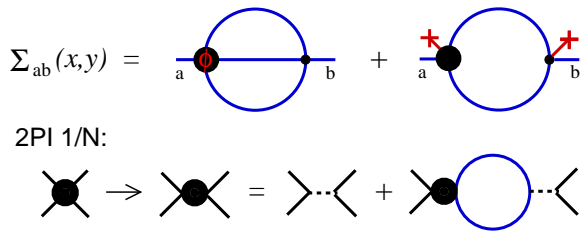}
}
\end{center}
\vspace*{-3ex}
\caption{
(Color online) 
Next-to-leading order (NLO) contributions to the scattering integral $J$.
Upper panel: Diagrammatic representation of the NLO contributions in the $1/\cal N$-expansion, to the self energy $\Sigma_{ab}(x,y;\phi)$.
The big filled circle represents a resummed vertex which in NLO $1/\mathcal{N}$ is defined as shown in the lower panel.
The (red) $\phi$ dependence of the vertex in the two-loop diagram contributing to $\Sigma$ indicates the further internal field dependence integrated over within the loops as can be seen from applying \Eq{SigmafromGamma2} to the lower line of diagrams of $\Gamma_{2}^{\mathrm{NLO}}$ in \Fig{DiagrExpGamma2NLO}.
All other symbols are as in \Fig{DiagrExpGamma2NLO}.
}
\label{fig:SigmaNLO1N}
\end{figure}

The NLO \OoN self energy entering $J(p)$, \Eq{qp-ansatz}, is depicted in \Fig{SigmaNLO1N}.
Before proceeding to the scaling behavior of $J$ we derive, in this section, the formal expression for $J$ in terms of the correlation functions $F$ and $\rho$.

\subsubsection{Vanishing field expectation value}
\label{sec:vanishingfield}
We first consider the simpler case of a vanishing field expectation value $\phi_a \equiv 0$. 
Scaling will result for strongly occupied modes where quantum fluctuations can be neglected.
Expressed in terms of the statistical and spectral components of $G$ this means that $F_{ab}(x)^2 \gg \rho_{ab}(x)^2$.
In thermal equilibrium this is readily obvious from the fluctuation-dissipation relation \eq{f-d-rel}.
As was shown in Refs.~\cite{Aarts:2001yn,Berges:2007ym} away from equilibrium the classical statistical limit implies that the $\rho\cdot\rho$ terms are neglected in the function $I^{F}$ defined in \Eq{IFrho} and, for $\phi\equiv0$, the $I^{\rho}\cdot\rho$ term in $\Sigma^{F}$, \Eq{SigmaNLO1N}.
The translationally invariant self-energies then read
\begin{align}
  \Sigma^F_{ab}(x)  
  &  =    -\lambda(I^F\cdot F_{ab})(x)
  \nonumber\\
  \Sigma^\rho_{ab}(x) 
  & =   -\lambda \big[ (I^\rho\cdot F_{ab})(x) + (I^F\cdot \rho_{ab})(x) \big],
\label{eq:Sigmarho}
\end{align}
where
\begin{equation}
  \lambda = 2g/\mathcal{N},
\end{equation}
and
\begin{align}
  I^F 
  & = 
  (1-  I^R) \ast \Pi^F
  +  I^F\ast \Pi^A,
  \nonumber\\
  I^\rho 
  & =   
  (1- I^R) \ast \Pi^\rho 
  +  I^\rho\ast \Pi^A,
\label{eq:Irhorecursive}
\end{align}
with the retarded and advanced functions
\begin{align}
  I^R(x) 
  & =  (\theta\cdot I^{\rho})(x),
  \quad
  I^A(x) 
   = (\theta^-\cdot I^{\rho})(x),
\label{eq:IA}
  \\
  \Pi^R(x) 
  & =  (\theta\cdot\Pi^{\rho})(x),
  \quad
  \Pi^A(x) 
   = (\theta^-\cdot\Pi^{\rho})(x).
\label{eq:PiA}
\end{align}
In the classical limit the functions $\Pi^{F,\rho}$ are
\begin{align}
  \Pi^F 
  & =  \lambda(F\cdot F)/2, 
 \quad
  \Pi^\rho 
   =  \lambda(F \cdot \rho). 
\label{eq:PiFrhoStatClass}
\end{align}
Here and in the following we use the more convenient notation in terms of products and convolutions of functions in $x$ as well as $p$ space, see \App{notation} for details.
The product/convolution of two correlation functions implies sums over field indices, $(F\cdot \rho)(x)=F_{ab}(x)\cdot\rho_{ab}(x)$.
Finite integration limits in time in the above convolutions are taken into account by the theta function $\theta(x)\equiv\theta(x_{0})$, with $\theta^{-}(x)\equiv\theta(-x)$.
Overall arguments $(x)$ have been suppressed. 

The recursive equations \eq{Irhorecursive} for $I^F$ and $I^\rho$ can be solved explicitly.
Using the convolution theorem one obtains, in momentum space,
\begin{align}
  I^F(p) 
  & =  (\lambda^{\mathrm{eff}}\cdot\Pi^{F})(p),
  \nonumber\\
  I^\rho(p) 
  & = (\lambda^{\mathrm{eff}}\cdot\Pi^{\rho})(p),
\label{eq:IF}
\end{align}
where $\Pi^{F}(p)=\lambda(F\ast F)(p)/2$, $\Pi^{\rho}(p)=\lambda(F\ast \rho)(p)$, and
\begin{align}
    \lambda^\mathrm{eff}(p)
    &=\frac{1-I^{R}}{1-\Pi^A} 
    = \frac{1}{(1-\Pi^A)(1+\Pi^R)} 
    \nonumber\\
    &= \frac{1}{\left|1+\Pi^R\right|^2}.
\label{eq:lambdaeff}
\end{align}
Here, we have suppressed arguments $(p)$.
The second equality in \Eq{lambdaeff} follows from 
\begin{equation}
  (1+\Pi^R(p))\cdot \left( 1-I^{R}(p) \right) = 1
\label{eq:corelambda}
\end{equation}
which in $x$-space reads:
\begin{equation}
  \Pi^R -I^{R} - \Pi^R\ast I^{R} = 0.
\end{equation}
This identity is proven by substituting the expression \eq{Irhorecursive} for $I^\rho$  into the second term giving
\begin{equation}
  \Pi^R + \theta\cdot\Big( \Pi^{\rho} - (\theta\cdot I^\rho)\ast \Pi^{\rho} + I^\rho\ast(\theta^-\cdot\Pi^{\rho})   \Big) 
  - \Pi^R\ast(\theta\cdot I^\rho) = 0
\end{equation}
which is verified using \eq{PiA} as well as that for any two functions $f$ and $g$:
$((\theta\cdot f)\ast(\theta\cdot g)) = \theta\cdot((\theta\cdot f)\ast(\theta\cdot g))$.

We are now ready to write down an expression for $J$ in the case of vanishing $\phi$: 
Substituting Eqs.~\eq{IF} into \eq{Sigmarho} and the self energies into condition \eq{stationarity}, we obtain for $J^0(p) \equiv J^{\phi=0}(p)$:
\begin{align}
  J^0(p) 
  &= -\frac{\lambda^{2}}{2}\Big\{
     F_{ba}\cdot
         \big[F_{ab} \ast \big(2\lambda^\mathrm{eff} \cdot(F\ast \rho)\big)
  \nonumber\\
     &\qquad\qquad\quad 
        +\ \rho_{ab}\ast \big(\lambda^\mathrm{eff}\cdot (F\ast F))\big] 
  \nonumber\\
     &\qquad\
     -\ \rho_{ba}\cdot
         \big[F_{ab}\ast \big(\lambda^\mathrm{eff}\cdot(F\ast F))\big]\Big\} 
     \nonumber\\
   =  -&\frac{\lambda^{2}}{2(2\pi)^8}\int\!\mathrm{d}^{d+1}k\,\mathrm{d}^{d+1}q\,
   \mathrm{d}^{d+1}r\, \delta(p+k-q-r)
    \nonumber\\
   &\!\!\!\times\ \lambda^\mathrm{eff}(p+k)\big[ 
     F_{ab}(p)F_{ab}(k)F_{cd}(q)\rho_{cd}(r) 
     \nonumber\\
  &\qquad\qquad\  \,
   +\ F_{ab}(p)F_{ab}(k)\rho_{cd}(q)F_{cd}(r) 
     \nonumber\\
  & \qquad\qquad\  \, 
    -\ F_{ab}(p)\rho_{ab}(k)F_{cd}(q)F_{cd}(r) 
     \nonumber\\
  & \qquad\qquad\  \, 
    -\ \rho_{ab}(p)F_{ab}(k)F_{cd}(q)F_{cd}(r)
     \big].
     \label{eq:J0}
\end{align}
%

\subsubsection{Non-vanishing field}
\label{sec:nonvanishingfield}
In the general case one needs to take into account the possibility of a non-vanishing condensate field $\phi\neq0$. 
As pointed out above, scaling will require that statistical fluctuations dominate over quantum fluctuations, $F_{ab}(x)^2 \gg \rho_{ab}(x)^2$.
In this case the self-energies \eq{SigmaNLO1N} read
\begin{align}
   \Sigma^F_{ab}(x) 
   & = \Sigma^{F(\phi=0)}_{ab}-\lambda\left( I^F\phi_a\phi_b + P^F\cdot F_{ab}\right),
   \label{eq:SigmaFfull} 
   \\
   \Sigma^\rho_{ab}(x) 
   & = \Sigma^{\rho(\phi=0)}_{ab} - \lambda\left(I^\rho\phi_a\phi_b + P^\rho\cdot F_{ab} + P^F\cdot\rho_{ab}\right).
\label{eq:Sigmarhofull}
\end{align}
Here and in the following we suppress arguments, e.g. $I^F=I^F(x)$, where they can be inferred.

$I^F$ and $I^\rho$ are as before, and $P^{F}$, $P^{\rho}$ are rewritten analogously as described in \Appendix{PiFrho}, giving
\begin{align}
  P^F(p) 
 & =  \lambda\phi_a\phi_b \big[F_{ab}\cdot\lambda^\mathrm{eff} - I^F\cdot\Delta_{ab}\big]
  \label{eq:PFDelta}
  \\
  P^\rho(p) 
 & =  \lambda\phi_a\phi_b\big[\rho_{ab}\cdot\lambda^\mathrm{eff} - I^\rho\cdot\Delta_{ab}\big], 
 \label{eq:PrhoDelta}
\end{align}
where (see \Appendix{PiFrho})
\begin{align}
   \Delta_{ab}(p) 
   &=  2\Re\left[ \frac{G^R_{ab}}{1+\Pi^R} \right] .
   \label{eq:Delta} 
\end{align}

We are now ready to derive the full function $J$. 
The ingredients $I^F$, $I^\rho$, $P^F$, and $P^\rho$ are given by Eqs.~\eq{IF}, \eq{PFDelta}, and \eq{PrhoDelta}, respectively. 
Substituting these into the full self-energies, \eq{SigmaFfull} and \eq{Sigmarhofull}, we get $J=J^{0}+J^{\phi}$ where $J^0$ was obtained in \Eq{J0} and the field dependent part reads
\begin{align}
  J^\phi 
  &= J^\lambda + J^\Lambda,
  \nonumber\\
  J^\lambda(p)
  &=  \frac{\lambda^{2}}{2}\phi_a\phi_b\cdot \big[ 
  \rho_{ab}\cdot(F\ast F)\cdot\lambda^\mathrm{eff} 
  \nonumber\\
  &\qquad\qquad\  +\ 2\rho_{cd}\cdot(F_{dc}\ast(F_{ab}\cdot\lambda^\mathrm{eff}))    
  \nonumber\\
  &\qquad\qquad\   -\ 2 F_{ab}\cdot(F\ast\rho)\cdot\lambda^\mathrm{eff} 
  \nonumber\\
  &\qquad\qquad\   -\ 2 F_{cd}\cdot(F_{dc}\ast(\rho_{ab}\cdot\lambda^\mathrm{eff})) 
  \nonumber\\
  &\qquad\qquad\   -\ 2F_{cd}\cdot(\rho_{dc}\ast(F_{ab}\cdot\lambda^\mathrm{eff})) 
  \big],
  \label{eq:JlambdaCompact}
  \\
  J^\Lambda(p)
  &= -\frac{\lambda^{3}}{2}\big(
  \rho_{ba}\cdot\{F_{ab}\ast [\Lambda\cdot(F\ast F)]\}  
  \nonumber\\
  &\qquad + 2 F_{ba}\cdot\{F_{ab}\ast[(F\ast \rho)\cdot\Lambda]\} 
  \nonumber\\
  &\qquad  + F_{ba}\cdot\{\rho_{ab}\ast[(F\ast F)\cdot\Lambda]\},
  \label{eq:JDeltaCompact}
\end{align}
with the effective coupling
\begin{equation}
  \Lambda(p) 
  = \phi_a\phi_b\Delta_{ab}\cdot\lambda^\mathrm{eff}.
\label{eq:Lambda}
\end{equation}
$J^{\lambda}$ and $J^\Lambda$ written in the form of momentum integrals as in \Eq{J0} for $J^{0}$ are provided in \App{PiFrho}. 
To make contact with the notation of Ref.~\cite{Berges:2008wm}, we introduce
\begin{equation}
  J^4 = J^0+J^\Lambda \quad\mathrm{and}\quad J^3 = J^\lambda.
  \label{eq:J4}
\end{equation}
The integral $J^{4}$ results from the left self-energy diagram in the upper panel of \Fig{SigmaNLO1N}, the integral $J^{3}$ from the right one in which the field $\phi$ is not integrated over inside the loops.
This concludes the calculation of $J$. We have three terms $J=J^0+J^\lambda+J^\Lambda$ given by Eqs.~\eq{J0}, \eq{Jlambda}, and \eq{JDelta}.

\section{Turbulent dynamical scaling}
\label{sec:TurbDynScaling}
%
\subsection{The scaling ansatz}
\label{sec:scalingansatz}
We are looking for solutions that fulfill the stationarity condition \eq{stationarity} in the infrared. Following Refs.~\cite{Berges:2008wm,Berges:2008sr} we consider scaling solutions
with properties
\begin{align}
  \rho_{ab}(s^zp_0,s\vec p) 
  & =  s^{-2+\eta}\rho_{ab}(p_0,\vec p) 
\label{eq:scalingrho} \\
  F_{ab}(s^zp_0,s\vec p) 
  & =  s^{-2-\kappa}F_{ab}(p_0,\vec p), \quad s>0 .
\label{eq:scalingF}
\end{align}
The free spectral function \eq{freerho} fulfills the scaling property \eq{scalingrho}, with $z=2$ and $\eta=0$. Deviations from this scaling are accounted for by a modified value for $z$ and an ``anomalous dimension''  $\eta$ (see e.g.~Ref.~\cite{Cardy1996a}) different from zero, which we will assume to be small in the following.
The scaling ansatz \eq{scalingF} for $F$ is chosen analogously, with an exponent $\kappa$ still to be determined. 

The scaling relation \eq{scalingrho}, with $\eta=0$, remains valid beyond the case of an ideal gas,  for the near-zero-temperature weakly interacting Bose gas in the broken phase which is described by Bogoliubov theory and merely a free gas of quasiparticles interacting with the condensate mode.
In this case, the dispersion becomes linear in $|\vec{p}|$ at low momentum, reflecting the sound-wave character of the modes and requiring $z=1$ in the scaling relations, but quadratic scaling is ensured by the Bogoliubov coefficients.
At zero temperature one has
\begin{align}
  &\rho_{ab}(p) 
  = i\pi\left[\delta(p_0 - \omega_{\vec p}) + \delta(p_0 + \omega_{\vec p})\right]
  \nonumber\\
  &\quad\times\ [ (u_{\vec p}^{2}+v_{\vec p}^{2})\delta_{ab}\,\mathrm{sgn}(p_0)-\sigma^{2}_{ab}],
\label{eq:rhoBogoliubov}
\end{align}
where $\omega_{\vec p}=[\epsilon_{\vec p}(\epsilon_{\vec p}+2g\phi_{a}\phi_{a})]^{1/2}$ (with $\epsilon_{\vec{p}}=\vec{p}^{2}/2m$) is the Bogoliubov dispersion which is linear in $|\vec{p}|$ at $\vec{p}^{2}\ll 4g \phi_{a}\phi_{a}$.
$u_{\vec p}=[(\epsilon_{\vec p}+g\phi_{a}\phi_{a}+\omega_{\vec p})/2\omega_{\vec p}]^{1/2}$ and $v_{\vec p}=[(\epsilon_{\vec p}+g\phi_{a}\phi_{a}-\omega_{\vec p})/2\omega_{\vec p}]^{1/2}$ are the Bogoliubov coefficients resulting from diagonalization of the propagator and scale like $u_{\vec p}\sim|{\vec p}|^{-1/2}$, $v_{\vec p}\sim|{\vec p}|^{-1/2}$ in the infrared.
Hence, the scaling \eq{scalingrho} also applies to function \eq{rhoBogoliubov}.

In the limit of large occupation numbers, the density of particles with momentum $\vec p$ in momentum space, $n(\vec p)=\langle\Psi^{\dagger}(\vec p)\Psi(\vec p)\rangle$, is given by 
\begin{align}
  n(\vec p) 
  = \frac{1}{2}\int\frac{\mathrm{d}\!p_0}{2\pi}\left[F_{11}(p) + iF_{12}(p)-i F_{21}(p) + F_{22}(p)\right].
\end{align}
Hence, the scaling behavior \eq{scalingF} of $F$ implies that
\begin{align}
  n(s\vec p) = s^{z-2-\kappa}n(\vec p).
  \label{eq:ParticleNoScaling}
\end{align}
Since in addition we assume isotropy, it follows that $n(\vec p) = n(|\vec p|) \sim |\vec p|^{z-2-\kappa}$. 
Hence, for a quadratic single-particle dispersion, $z=2$, $\kappa$ acts as the occupation-number scaling exponent.

In order to derive the scaling exponent $\kappa$ from the necessary condition \eq{stationarity} for stationarity we take into account the spatial isotropy of the Hamiltonian  and consider the ensuing weaker necessary condition which results from integrating \Eq{stationarity} over the spatial momenta $\vec p$ \footnote{%
Alternatively one could consider the condition that the function obtained by integrating $J(p)$ over the frequency $p_{0}$ and the $d-1$-dimensional angular dependence of $\vec p$, leaving a function of $|\vec p|$.
The scaling analysis of such a function would, however, yield the same exponents as obtained from \Eq{weakstationarityp0}.}, 
\begin{align}
  J(p_{0}) 
  &\equiv \int \ddf{p} \left[\Sigma^\rho_{ab}(p)F_{ba}(p) - \Sigma^F_{ab}(p)\rho_{ba}(p)\right] \stackrel{!}{=} 0.
\label{eq:weakstationarityp0}
\end{align}

In the following we determine, under the assumption that the correlation functions obey the above scaling properties, for which values of $\kappa$ the stationarity condition \eq{weakstationarityp0} is fulfilled.  
To derive these values we make use of scaling transformations~\cite{Berges:2008wm,Berges:2008sr} which imply reparametrizations of the frequency-momentum integrals in $J(p_{0})$, similar to the Zakharov transformations used in the context of weak wave turbulence \cite{Zakharov1992a}. 
The thus reparametrized integrals can be transformed to a unique form with the help of the scaling laws for the correlation functions which relate these functions at different frequency and momentum scales to each other.

In the following we will consider the scaling behavior in the ultraviolet (UV) and infrared (IR) regimes separately.
We first derive the scaling exponents  
in the UV regime. They correspond the perturbative Kolmogorov exponents, $\kappa_{P}$, \Eq{KolPScaling}, and $\kappa_{Q}$, \Eq{KolQScaling}. 
We will then present the derivation of a different scaling regime, with significantly enlarged values for exponents, in the IR.

\subsection{Scaling exponents in the ultraviolet regime}
\label{sec:kappaUV}

The aim of this section is to show that in the perturbative regime of sufficiently large frequencies and momenta the exponents characterizing weak wave turbulence
\begin{equation}
\kappa = d+ z -\frac{8}{3} + \frac{\eta}{3},
  \label{eq:kappaQJ0p0UVfinal}
\end{equation}
and
\begin{equation}
  \kappa = d   + \frac{4}{3}(z-2)+ \frac{\eta}{3}
  \label{eq:kappaUVfinal}
\end{equation}
allow $J(p_{0})$, \Eq{weakstationarityp0}, in $d$ spatial dimensions to vanish. Here \Eq{kappaQJ0p0UVfinal} corresponds to \Eq{KolQScaling} and \Eq{kappaUVfinal} to \Eq{KolPScaling} for $z=2$ and $\eta = 0$. 
The conditions for the existence of these scaling solutions are different.
The solution with $\kappa$ given by \Eq{kappaQJ0p0UVfinal} requires that the excitations of the system are described sufficiently well by quasiparticles,  \Eq{qp-ansatz}, with a fixed dispersion relation  \eq{freerho} while the solution \eq{kappaUVfinal} exists also without this restriction.
In the UV the leading-order perturbative approximation with a zero-width spectral function becomes valid such that kinetic theory and therefore both scaling solutions apply.

\subsubsection{Dominant contribution to $J$}
\label{sec:dominantUV}
$J$ has three components $J = J^0 + J^\lambda + J^\Lambda$, given by Eqs.~\eq{J0}, \eq{Jlambda}, and \eq{JDelta}. 
We derive scaling laws for $J^0$, $J^\lambda$ and $J^\Lambda$ in the UV limit to determine which of the contributions to $J(p)$ dominates in this regime. For sufficiently large frequencies and momenta the occupation numbers are low. As a consequence, one can classify the contributions to \Eq{weakstationarityp0} perturbatively according to powers of $\lambda$. Since any non-zero width of the spectral function is of higher order in $\lambda$, a $\delta$-like spectral function can be assumed under the integral of \Eq{weakstationarityp0} to a given order in the coupling. A similar reasoning applies to the perturbative statistical function. Therefore, we will recover the results of kinetic theory.   

Turning first to $J^0$ we need to find the scaling law for the effective coupling $\lambda^\mathrm{eff}$ defined in \Eq{lambdaeff}.
As discussed in Appendix \ref{app:ScalingProperties}, $\Pi^R$ scales like
$\Pi^R(s^zp_0,s\vec p) = s^{z+d-4-\kappa+\eta}\Pi^R(p_0,\vec p)$
(see \Eq{PiRscalingAppendix}). 
If $\kappa > z+d-4+\eta$ which according to \Eq{kappaUVfinal} requires $\eta<z/2+2$, this implies that $|\Pi^R(p)| \ll 1$ in the high-momentum limit. 
Hence, one can neglect $\Pi^{R}$ in the denominator of \eq{lambdaeff}, and obtains:
\begin{equation}
  \lambda^\mathrm{eff}(p) \approx 1 \qquad \textrm{(UV limit)}.
\end{equation}
Away from the UV limit the above relation becomes approximative as does the scaling behavior discussed in the following.
This also applies to the kinetic wave-turbulence theory where one needs to show the locality of the scattering integral.
Locality implies, that the scaling law at a particular momentum scale is independent of what happens far away in momentum space.
In kinetic theory, the locality of the scattering integral follows from the convergence of the integral \cite{Zakharov1992a}.
We note that a proof of convergence of the scattering integral in the full dynamical theory is more difficult than in the kinetic approximation, where the dispersion is fixed and the frequency integrals can be eliminated.
The existence of approximate scaling solutions at finite momenta can be shown by means of numerical simulations of the dynamics as presented, e.g., in Refs.~\cite{Micha:2004bv,Berges:2008wm}.

It follows that in the UV limit, we can neglect the effective coupling in $J^{0}(p)$, \Eq{J0}.
The same arguments apply to the remaining contributions $ J^\lambda$ and $J^\Lambda$, and inserting the scaling properties \eq{scalingrho}, \eq{scalingF}, \eq{scalinglambdaeffUV}, \eq{scalingDeltaUV}, and \eq{scalingLambdaUV} into Eqs.~\eq{J0}, \eq{Jlambda}, and \eq{JDelta} one finds the scaling relations
\begin{align}
  J^0(s^zp_0,s\vec p) 
  & =  s^{2(d+z-4)+\eta-3\kappa} J^0(p_0,\vec p),
  \\
  J^\lambda(s^zp_0,s\vec p) 
  & =  s^{d+z-6+\eta-2\kappa} J^\lambda(p_0,\vec p),
  \\
  J^\Lambda(s^zp_0, s\vec p)  
  & = s^{2(d+z-5)+2\eta-3\kappa} J^\Lambda(p_0,\vec p).
\end{align}

We anticipate the general result $\kappa=d+4(z-2)/3+\eta/3$.
Comparing the scaling laws for $J^0$, $J^\lambda$ and $J^\Lambda$, which scale with exponents $2(d+z-4)+\eta-3\kappa=-d-2z$, $d+z-6+\eta-2\kappa=-d-(5z+2+\eta)/3$ and $2(d+z-5)+2\eta-3\kappa=-d-2z-2+\eta$, we come to the conclusion that both $J^0$ and $J^\lambda$ dominate in the limit of large momenta for $z=2$, assuming $\eta=0$, while for $z<2$ or $\eta>0$ the integral $J^{0}$, and for $z>2$, the integral $J^{\lambda}$ dominate on their own. 
It will turn out that the same value of $\kappa=d+\eta/3$ renders $J^0$ and $J^\lambda$ to vanish simultaneously if $z=2$.%

\subsubsection{Scaling transformations}
Taking into account the possible dominant contributions in the UV regime (see last paragraph) the stationarity condition \eq{weakstationarityp0} for the momentum integral over $J(p)$ requires 
\begin{equation}
  J^0(p_0) \stackrel{!}{=} 0.
  \label{eq:conditionJ0lambdaUV}
\end{equation}
We derive a relation between $\kappa$, $\eta$, and $z$ which fulfills this condition. 
Using the above argument that we can set $\lambda^{\mathrm{eff}}(p+k)=1$ in the integrand, the integral reads 
\begin{align}
  J^0(p_{0}) 
  & =  -\frac{\lambda^{2}}{2(2\pi)^8}\int\ddf{p}\int\mathrm{d}^{d+1}k\,
  \mathrm{d}^{d+1}q\, \mathrm{d}^{d+1} r\, 
    \nonumber\\
  &\qquad \qquad
   \times\,
  \delta(p+k-q-r)
    \nonumber\\
  &\qquad \qquad
   \times\ \big[ 
     F_{ab}(p)F_{ab}(k)F_{cd}(q)\rho_{cd}(r) 
     \nonumber\\
  &\qquad\qquad\  \,
   +\ F_{ab}(p)F_{ab}(k)\rho_{cd}(q)F_{cd}(r) 
     \nonumber\\
  & \qquad\qquad\  \, 
    -\ F_{ab}(p)\rho_{ab}(k)F_{cd}(q)F_{cd}(r) 
     \nonumber\\
  & \qquad\qquad\  \, 
    -\ \rho_{ab}(p)F_{ab}(k)F_{cd}(q)F_{cd}(r)
     \big].
     \label{eq:J0UV}
\end{align}
As is described in more detail in App.~\ref{app:Zakharov}, it is possible with the help of scaling transformations to exchange $p$ with another integration variable in \Eq{J0UV}, thereby keeping $p_0$ as a free variable \cite{Berges:2008wm,Berges:2008sr}.
These transformations are similar to the Zakharov transformations typically employed to compute exponents for weak wave turbulence for a fixed dispersion relation and $z=2$, $\eta = 0$. In that case, frequency  integrals are separated into different domains of integration, which can be mapped onto each other~\cite{Zakharov1992a}. Here we use scaling transformations to map the different integrands of the integral \Eq{J0UV}. Since we do not assume a fixed dispersion relation in this approach, we will be able to use the same method also to compute IR scaling exponents in the non-perturbative regime below.   
This transformation makes use of the scaling properties \eq{scalingrho} and \eq{scalingF}. 
As these scaling relations involve an $s^z$ in the $p_0$ component, one first rewrites \Eq{J0UV} such that the integration variables $k_0$, $q_0$, and $r_0$ are positive:
\begin{align}
  &\int\mathrm{d}k_0\mathrm{d}q_0\mathrm{d}r_0\ f(p,k,q,r)
   =  \int\limits_{k_0>0,q_0>0,r_0>0} \mathrm{d}k_0\mathrm{d}q_0\mathrm{d}r_0
  \nonumber\\
  &\quad\times\ 
  \big( f(p,k,q,r) + [r\to -r] + [q \to -q] + [k \to -k] 
  \nonumber \\
  &\qquad\quad  +\ 
  [q \to -q,r\to -r] + [k \to -k,r\to -r] 
  \nonumber \\
  &\qquad \quad  +\ 
  [k \to\! -k,q \to -q]  
  \nonumber \\
  &\qquad \quad  +\ 
   [k \to\! -k,q \to -q,r\to -r] \big).
\end{align}
Here, the function $f$ contains the full integrand in \Eq{J0UV}, including the remaining spatial integrals.
For $p_{0}>0$, the fifth summand in parantheses vanishes due to the delta-function $\delta(p+k-q-r)$, such that there are $4\cdot7=28$ terms contributing to the integrand in \Eq{J0UV}. 
Negative arguments can easily be dispelled using the symmetry properties for $F$ and $\rho$:
${F}_{ab}(-p) = {F}_{ba}(p)$, ${\rho}_{ab}(-p) = -{\rho}_{ba}(p)$.

At this point we use a scaling transformation (see App.~\ref{app:Zakharov}) to achieve that in each of the 28 summands, $\rho$ carries the $p$-argument. 
We furthermore use that we can permute the integration variables $k$, $q$, $r$ at will.
Combining terms that are equivalent one arrives at the final result:
\begin{align}
  &J^0(p_0) 
  =  \int\limits_{k_0>0,q_0>0,r_0>0}   \ddf{p}
  \int\mathrm{d}^{d+1}k\,\mathrm{d}^{d+1}q\,\mathrm{d}^{d+1}r
  \nonumber\\
  &\quad\times\ \big( 
  \delta_{p+k-q-r} \stackrel{ab}{\rho}_p\stackrel{ab}{F}_k\stackrel{cd}{F}_q\stackrel{cd}{F}_r
     p_0^{-\Delta}(p_0^\Delta + k_0^\Delta - q_0^\Delta - r_0^\Delta )   
     \nonumber \\
  &\quad +\
  \phantom{2} \delta_{p-k+q+r} \stackrel{ab}{\rho}_p\stackrel{ba}{F}_k\stackrel{cd}{F}_q\stackrel{cd}{F}_r
     p_0^{-\Delta}(p_0^\Delta - k_0^\Delta + q_0^\Delta + r_0^\Delta )  
     \nonumber \\
  &\quad +\
  \phantom{2} \delta_{p-k-q-r} \stackrel{ab}{\rho}_p\stackrel{ba}{F}_k\stackrel{cd}{F}_q\stackrel{cd}{F}_r
     p_0^{-\Delta}(p_0^\Delta - k_0^\Delta - q_0^\Delta - r_0^\Delta )
    \nonumber \\
  &\quad +\
  2 \delta_{p+k-q+r} \stackrel{ab}{\rho}_p\stackrel{ab}{F}_k\stackrel{cd}{F}_q\stackrel{dc}{F}_r
     p_0^{-\Delta}(p_0^\Delta + k_0^\Delta - q_0^\Delta + r_0^\Delta )    
     \nonumber \\
  &\quad +\
  2 \delta_{p-k-q+r} \stackrel{ab}{\rho}_p\stackrel{ba}{F}_k\stackrel{cd}{F}_q\stackrel{dc}{F}_r
     p_0^{-\Delta}(p_0^\Delta - k_0^\Delta - q_0^\Delta + r_0^\Delta ) 
  \big),
\label{eq:J0UVfinal} 
\end{align}
Here we have used the short-hand notation $\stackrel{ab}{\rho}_p = \rho_{ab}(p)$, cf.~\Eq{rhoStackedIndices}, and $\delta_{p}=\delta(p)$.
Combining the scaling of correlation functions, delta function, and integral measures, the exponent $\Delta$ results as
\begin{equation}
  \Delta = \frac{1}{z}(3\kappa - 3d + 8 - \eta) - 3.
  \label{eq:betaJ0UV}
\end{equation}

We now consider the condition that $J^{\lambda}(p_{0})$, \Eq{Jlambda} vanishes,
\begin{equation}
  J^\lambda(p_0) \stackrel{!}{=} 0.
  \label{eq:conditionJlambdaUV}
\end{equation}
As before, if $p_0$ is chosen large, only the high-momentum behavior of the functions involved is important, in particular $\lambda^\mathrm{eff}\approx 1$.
Rewriting the integral such that all frequency variables are positive and neglecting terms vanishing by the delta function, one obtains $9\cdot3=27$ summands. 
By use of a scaling transformation of the form defined in App.~\ref{app:Zakharov}, \Eq{Zakharovp0}, each of the 27 terms is rewritten such that $\rho$ carries the $p$ argument. 
Analogous steps as before lead to the final expression
\begin{align}
  &J^\lambda(p_0) 
   =  \phi_a\phi_b\int\limits_{k_0>0,q_0>0}   \ddf{p}\int\mathrm{d}^{d+1}k\,\mathrm{d}^{d+1}q\,
  \nonumber\\
  &\quad\times\ \big( 
  \delta_{p-k-q}\stackrel{ab}{\rho}_p\stackrel{cd}{F}_k\stackrel{cd}{F}_q \cdot p_0^{-\Delta}(p_0^\Delta-k_0^\Delta-q_0^\Delta)      
     \nonumber \\
  &\quad +\
 \phantom{2} \delta_{p-k+q}\stackrel{ab}{\rho}_p\stackrel{cd}{F}_k\stackrel{dc}{F}_q \cdot p_0^{-\Delta}(p_0^\Delta-k_0^\Delta+q_0^\Delta)      
     \nonumber \\
  &\quad +\
 \phantom{2} \delta_{p+k-q}\stackrel{ab}{\rho}_p\stackrel{cd}{F}_k\stackrel{dc}{F}_q \cdot p_0^{-\Delta}(p_0^\Delta+k_0^\Delta-q_0^\Delta)      
     \nonumber \\
  &\quad +\
 2\delta_{p-k-q}\stackrel{cd}{\rho}_p\stackrel{dc}{F}_k\stackrel{ab}{F}_q \cdot p_0^{-\Delta}(p_0^\Delta-k_0^\Delta-q_0^\Delta)       
     \nonumber \\
  &\quad +\
 2\delta_{p-k+q}\stackrel{cd}{\rho}_p\stackrel{dc}{F}_k\stackrel{ab}{F}_q \cdot p_0^{-\Delta}(p_0^\Delta-k_0^\Delta+q_0^\Delta)       
     \nonumber \\
  &\quad +\
 2\delta_{p+k-q}\stackrel{cd}{\rho}_p\stackrel{cd}{F}_k\stackrel{ab}{F}_q \cdot p_0^{-\Delta}(p_0^\Delta+k_0^\Delta-q_0^\Delta) 
 \big),
\label{eq:JlambdaUVfinal} 
\end{align}
with
\begin{equation}
  \Delta = \frac{1}{z}(2\kappa-2d+6-\eta)  - 2.
\end{equation}
%

\subsubsection{Scaling exponent: Results in the UV regime}
\label{sec:kappaUVresults}
The delta distributions ensure that $J^0(p_{0})$, \Eq{J0UVfinal}, vanishes if
\begin{equation}
\Delta=1 \quad \Leftrightarrow \quad \kappa=d+\frac{4}{3}(z-2)+\frac{\eta}{3}.
  \label{eq:kappaJ0p0UV}
\end{equation}
$J^\lambda(p_{0})$, \Eq{JlambdaUVfinal}, becomes manifestly zero if
\begin{equation}
  \Delta=1 \quad \Leftrightarrow \quad \kappa=d +\frac{3}{2}(z-2)+\frac{\eta}{2}.
  \label{eq:kappaJlambdap0UV}
\end{equation}
In the special case of a vanishing field expectation value, $\varphi=0$, $J=J^0$, and the result for $\kappa$ remains unaffected.
We therefore come to the conclusion that in a cold Bose gas with quadratic dispersion, $z=2$, and $\eta=0$ Kolmogorov-like scaling occurs in the UV regime, with $\kappa=d$.
The exponent \eq{kappaUVfinal} implies a scaling of the particle number, \Eq{ParticleNoScaling}, as
\begin{align}
  n(|\vec p|) \sim |\vec p|^{-d-(z-2+\eta)/3}.
  \label{eq:UVParticleNoScaling}
\end{align}

The scaling exponent $\kappa=d$ obtained for $z=2$ and $\eta=0$ corresponds to the Kolmogorov exponent $\kappa_{P}$, \Eq{KolPScaling}, for weak wave turbulence in the kinetic approximation, see \Sect{KinFP}. We now consider the derivation of $\kappa_{Q}$, \Eq{KolQScaling}.

Inserting the perturbative quasiparticle behavior, \Eq{qp-ansatz}, into \Eq{J0UVfinal} causes the first four lines in round parentheses to vanish because, after evaluating the frequency integrals over $k_{0}$, $q_{0}$, and $r_{0}$, and thus setting the frequencies to the respective quasiparticle frequencies, $q_{0}=\omega(\vec q)$, etc., one finds that 
\begin{align}
  {\rho}_{ab}(\omega_{\vec k},{\vec k}){F}_{ab}(\omega_{\vec k'},{\vec k'}) 
   = {F}_{ab}(\omega_{\vec k},{\vec k}){F}_{ab}(\omega_{\vec k'},{\vec k'}) = 0,
   \label{eq:vanishingFabFab}
\end{align}
for arbitrary momenta $\vec k$, $\vec k'$, while in general ${\rho}_{ab}(\omega_{\vec k},{\vec k}){F}_{ba}(\omega_{\vec k'},{\vec k'})\not=0$, ${F}_{ab}(\omega_{\vec k},{\vec k}){F}_{ba}(\omega_{\vec k'},{\vec k'})\not=0$.
As a consequence, only the last term $\propto p_0^\Delta - k_0^\Delta - q_0^\Delta + r_0^\Delta$ remains in \Eq{J0UVfinal}, and $J^{\lambda}(\omega_{\vec p})$ vanishes identically, cf.~\Eq{JlambdaUVfinal}.
Hence, a further scaling exponent results since $J^{0}(p_{0})$ vanishes now also for
\begin{equation}
\Delta=0 \quad \Leftrightarrow \quad \kappa=d+ z -\frac{8}{3}+\frac{\eta}{3}.
  \label{eq:kappaQJ0p0UV}
\end{equation}
For $z=2$ and $\eta=0$ we obtain $\kappa=d-2/3$.
We emphasize that a fixed dispersion relation or \eq{qp-ansatz} is required to find scaling with this exponent.
Without \eq{qp-ansatz}, $\Delta=1$ leaves in general nonvanishing contributions to $J^{0}(p_{0})$, \Eq{J0UVfinal}, as well as to $J^{\lambda}(p_{0})$, \Eq{JlambdaUVfinal}.
These terms are in general nonzero since the frequency dependences of $F$ and $\rho$ allow for collision events which, as is read off the delta functions in, e.g., \Eq{J0UVfinal}, change the number of wave excitations.
Hence, the solution $\kappa_{Q}$, \Eq{KolPScaling} only holds in the quasiparticle limit, \Eq{qp-ansatz}.
This finding is consistent with the fact mentioned in \Sect{KinFP} that Kolmogorov's exponent $\kappa_{Q}$ results under the assumption of a $\vec p$-independent radial quasiparticle flux $Q(|\vec p|)$ which in turn requires the existence of a well-defined quasiparticle number.

\subsection{Scaling exponents in the infrared regime}
\label{sec:kappaIR}
In this section we show that in the non-perturbative regime of small frequencies and momenta a different scaling behavior arises than what is found above for weak wave turbulence. The exponents
\begin{equation}
  \kappa = d+z-\eta
  \label{eq:kappaIRQ}
\end{equation}
or
\begin{equation}
  \kappa = d+2z-\eta
  \label{eq:kappaIRfinal}
\end{equation}
allow $J(p_{0})$, \Eq{weakstationarityp0} to vanish in the infrared. 
Despite the fact that there are strong corrections to kinetic theory in this regime, we point out that certain properties of the non-perturbative solutions \Eq{kappaIRQ} and \Eq{kappaIRfinal} are still similar to the Kolmogorov solutions \Eq{kappaQJ0p0UVfinal} and \Eq{kappaUVfinal}, respectively.
Major differences concern, apart from the numerical values, properties such as critical slowing down or universality for the infrared behavior, which is characteristic for critical phenomena, here far from equilibrium. Universality of a critical phenomenon requires that the integral equations, which determine the critical exponents \Eq{kappaIRQ} and \Eq{kappaIRfinal}, are dominated by the low-frequency and -momentum limit such that higher momenta do not affect the value of exponents. We also emphasize that, while the perturbative UV solutions are only valid at high momenta if quantum corrections are neglected, the non-perturbative IR solutions are a property of the quantum theory. Stated differently, classical-statistical fluctuations dominate over quantum corrections at low momenta, where occupations numbers are high, such that the quantum and the classical theory are characterized by exactly the same exponents.

\subsubsection{Dominant contribution to $J$}
$J$ has three components $J = J^0 + J^\lambda + J^\Lambda$, given by Eqs.~\eq{J0}, \eq{Jlambda}, and \eq{JDelta}. 
Again, we first derive scaling laws for $J^0$, $J^\lambda$ and $J^\Lambda$ in the IR limit to determine which of the contributions to $J(p)$ dominates in this regime.
Turning first to $J^0$ we need to find the scaling behavior of the effective coupling $\lambda^\mathrm{eff}$.
As discussed in Appendix \ref{app:ScalingProperties}, if $\kappa > d+z-4+\eta$, which according to \Eq{kappaIRfinal} requires $\eta<z/2+2$, then $|\Pi^R(p)| \gg 1$ in the denominator of $\lambda^ \mathrm{eff}$, \Eq{lambdaeff}, in the low-momentum limit.
Hence, $\lambda^\mathrm{eff}$, in the IR limit, scales like
\begin{align}
  \lambda^\mathrm{eff}(s^z p_0, s\vec p) 
  &= s^{2(\kappa+4-z-d-\eta)}\lambda^\mathrm{eff}(p_0,\vec p)
  \nonumber\\
  &= s^{2(z+4-2\eta)}\lambda^\mathrm{eff}(p_0,\vec p).
\label{eq:lambdaeffscaling}
\end{align}
In the last line we anticipated the result \eq{kappaIRfinal}.
Now consider $J^0(p)$ in the low-momentum range. 
Proceeding through the same steps as for the UV case one finds the IR scaling of the three integrals contributing to $J(p)$:
\begin{align}
  J^0(s^zp_0,s\vec p) 
  & =  s^{-\kappa-\eta} J^0(p_0,\vec p),
\label{J0scalingIR}
  \\
  J^\lambda(s^zp_0,s\vec p) 
  & =  s^{-d-z+2-\eta} J^\lambda(p_0,\vec p),
\label{JlambdascalingIR}
  \\
  J^\Lambda(s^zp_0, s\vec p)  
  & = s^{-d-z+2-\eta} J^\Lambda(p_0,\vec p).
\label{JDeltascalingIR}
\end{align}
Comparing these scaling relations we conclude that for $\kappa > d+z-2$, the integral $J^0$ dominates in the IR limit.
Given the result \eq{kappaIRfinal} this condition requires $\eta<z+2$.

\subsubsection{Scaling transformations}
Taking into account the dominant contributions in the IR regime (see last paragraph) the stationarity condition \eq{weakstationarityp0} reads
\begin{equation}
  J^0(p_0) \stackrel{!}{=} 0.
  \label{eq:conditionJ0lambdaIR}
\end{equation}
In the following we show that this leads to a different value for $\kappa$ as compared to the UV regime. 
We obtain, for $J^0(p_0)$, an expression identical to \Eq{J0UV}, except that an overall factor $\lambda^\mathrm{eff}({p+k})$ multiplies the integrand, cf.~\Eq{J0}. 
It is a remarkable property that the main consequence in the non-perturbative regime is the appearance of a momentum-dependent effective coupling. As a consequence, in this sense there is a direct link between the weak-wave-turbulence analysis in the UV and the calculation of exponents in the infrared regime of what may be called strong turbulence. 

We proceed like in the UV case, splitting the integral such that it contains only positive frequencies, applying scaling transformations of the type \eq{Zakharovp0} and finally combining terms. 
We obtain a result similar to \eq{J0UVfinal}, but a different expression for $\Delta$: 
As $\lambda^\mathrm{eff}$ now contributes a scaling factor this results as (compare \Eq{betaJ0UV}):
\begin{equation}
  \Delta = \frac{1}{z}(\kappa + \eta - d - z)
  \label{eq:IRbeta}
\end{equation}
Hence, $J^0(p_{0})$ vanishes identically if
\begin{equation}
  \Delta=1 \quad \Leftrightarrow \quad \kappa=d+2z-\eta.
  \label{eq:kappaJ0p0IR}
\end{equation}
This result also holds for the special case of vanishing field, $\phi = 0$, where $J=J^0$.
The exponent \eq{kappaIRfinal} implies a scaling of the particle number, \Eq{ParticleNoScaling}, as
\begin{align}
  n(|\vec p|) \sim |\vec p|^{-d-z-2+\eta}.
  \label{eq:NonpertIRParticleNoScaling}
\end{align}

A second solution is obtained from \Eq{IRbeta} for $\Delta=0$. 
\begin{equation}
  \Delta=0 \quad \Leftrightarrow \quad \kappa=d+z-\eta.
  \label{eq:kappaJ0p0IRQ}
\end{equation}
Given our discussion in \Sect{kappaUVresults}, this solution requires a fixed dispersion relation. While in the UV this is described in terms of a perturbative, i.e.\ $\delta$-like spectral function, this is not the case for the infrared properties of critical phenomena. 
Here the power-law behavior of the spectral function leads to dominant contributions in the infrared, i.e.\ for both $p^0$ and $|\vec{p}|$ approaching zero. For a detailed discussion of 
this in the context of equilibrium critical phenomena, see Ref.~\cite{Berges:2009jz}.  
In Refs.~\cite{Micha:2002ey,Micha:2004bv} numerical simulations of the classical equations of motion for a relativistic scalar theory with quartic self-interactions were presented which demonstrated the evolution of the system into a turbulent scaling regime after parametric resonance, confirming perturbative results for the exponents of weak wave turbulence. New numerical simulations were presented
in Ref.~\cite{Berges:2008wm}, which extend to the infrared and demonstrate the presence of a strong turbulence regime with strongly enhanced correlations. Using the 2PI effective action in NLO \OoN approximation, the properties of this scaling behavior in the infrared regime are recovered  analytically \cite{Berges:2008wm}.
The simulations indicate that after parametric resonance or spinodal decomposition dynamics, the infrared scaling exponent $\kappa=d+z-\eta$ is approached taking $z=1$ and $\eta = 0$. 
We emphasize that this is the exponent which requires a fixed dispersion relation as it is not expected in the IR regime.
Hence the numerical results of Ref.~\cite{Berges:2008wm}, together with our results indicate that the finite width of the dominant peak of the spectral function in the IR limit $p\to0$ allows for a scaling solution reminiscent of turbulence, corresponding to the Kolmogorov scaling with $\kappa_{Q}$ in the UV.
Moreover, according to the numerical results this should be approximately valid and representing a form of ``strong turbulence'' at small finite momenta.

\subsection{Thermal scaling}
\label{sec:thermal}
We close our analysis by considering the scaling at the thermal fixed point.
The obtained exponents $\kappa=d+z-\eta$ and $\kappa=d+z-8/3+\eta/3$ in the IR and UV limits, respectively, are larger than the exponent for thermal equilibrium: 
From the fluctuation-dissipation relation \eq{f-d-rel} and \eq{nBE}, it follows that, in the Raleigh-Jeans limit $n_\mathrm{BE} \gg 1$, i.e.\ $\beta\omega \approx 0$ and hence $e^{\beta\omega} \approx 1 + \beta\omega$, that
\begin{equation}
  F_{ab}(\omega,\vec p) \sim (\beta\omega)^{-1}\cdot\rho_{ab}(\omega,\vec p)
\end{equation}
and hence
\begin{equation}
F_{ab}(s^z\omega,s\vec p) = s^{-2-z+\eta}F_{ab}(\omega,\vec p).
\end{equation}
As a consequence, the thermal scaling exponent in the large-momentum limit reads
\begin{equation}
  \kappa = z-\eta  \quad \textrm{(thermal equilibrium)}.
\end{equation}
Note that in a Bose gas in the broken, i.e., condensate phase, at low momenta $\kappa$ is modified by the interactions.
Using the Bogoliubov dispersion $\omega_{\vec p}=[\epsilon_{\vec p}(\epsilon_{\vec p}+2g\phi_{a}\phi_{a})]^{1/2}$, with $\epsilon_{\vec{p}}=\vec{p}^{2}/2m$, which becomes linear in $|\vec{p}|$ at $\vec{p}^{2}\ll 4g \phi_{a}\phi_{a}$ one finds that, at zero temperature and with $\eta=0$,
\begin{equation}
  \kappa = 1 \quad (\textrm{IR limit},T=0).
\end{equation}
Taking into account  \Eq{ParticleNoScaling}, $\kappa=z-\eta$ leaves, however, the scaling of the particle number invariant as compared to the UV, scaling,
\begin{align}
  n(|\vec p|) \sim |\vec p|^{-2+\eta}.
  \label{eq:ThermalParticleNoScaling}
\end{align}
We note that it was shown in Ref.~\cite{Hohenberg1965a}, using sum rules for moments derived from linear response theory that $2n_{\vec p}+1 \ge 2k_{B}Tm\phi_{a}\phi_{a}/p^{2}$, which implies that, at finite temperature, $2-z+\kappa \ge 2$, i.e.,
\begin{equation}
  \kappa \ge z \quad (\textrm{IR limit},T>0).
\end{equation}
in the infrared limit.

\section{Conclusions}
\label{sec:Discussion}
We have presented a scaling analysis of wave-turbulent fixed points in the dynamical evolution of a spatially uniform ultracold Bose gas away from thermal equilibrium.
The analysis was focused on the scattering (self-energy) term of the dynamical equations for two-point correlation functions, in nonperturbative next-to-leading-order approximation of a $1/\mathcal{N}$ approximation.
The dynamic equations in this approximation were derived from the 2PI effective action which ensure the approximated time evolution to obey crucial conservation laws as those for total energy and particle number.

Searching for a dynamical fixed point was performed by implementing the necessary condition that the momentum integral over the scattering term vanishes in a nonthermal stationary state.
Presupposing scaling properties for the correlation functions entering this condition, scaling exponents were derived for the two-point functions characterizing a nonthermal stationary state.
Such a state, showing nontrivial scaling properties, is well known in the theory of wave turbulence.
We reproduce the perturbative Kolmogorov scaling properties discussed, e.g., in Ref.~\cite{Zakharov1992a},
identifying it as the scaling behavior in the high-frequency regime.
We find that the turbulent spectrum, i.e., the momentum mode occupation, in $d$ spatial dimensions, behaves like $|\vec{p}|^{-d-(z-2+\eta)/3}$ in this regime, where $d$ is the number of spatial dimensions, $|\vec{p}|^z$ the scaling of the dispersion, and $\eta$ an anomalous dimension assumed to be small. 
In deriving this scaling, a possible nonzero constant condensate field was taken into account, but the two-to-two scattering events independent of this field were found to dominate the scaling. 

As in the kinetic theory of weak wave turbulence, two different solutions are found in both the ultraviolet and infrared limits. 
Solutions which correspond to a quasiparticle cascade in the ultraviolet limit require a fixed dispersion relation between frequency and momentum while it is sufficient to have energy conservation for the set of solutions which relate to a scale-invariant energy flux, i.e., an energy cascade.
In the UV the leading-order perturbative approximation with a zero-width spectral function becomes valid such that kinetic theory and therefore both scaling solutions apply.
At momentum scales approaching zero the divergence in the spectral function dominates its behavior such that also in the IR limit both scaling solutions are relevant.

At low momenta and frequencies we find strong corrections to the kinetic theory underlying the perturbative Kolmogorov scaling analysis. A different scaling regime appears in the infrared, with significantly enhanced scaling exponents resulting from the nonperturbative character of the interactions in the gas. Here the nonperturbative nature of the 2PI $1/\mathcal{N}$ expansion to NLO is essential to be able to describe this physics. Our findings confirm analogous results presented for the relativistic case in Refs.~\cite{Berges:2008wm,Berges:2008sr}.
This scaling enhancement should show up in a strong occupation of low-momentum modes, rising with a power law $n(\vec p)\sim |\vec p|^{-d-z-2+\eta}$, as compared to the above quoted Kolmogorov scaling for high momentum modes. Since the phenomenon occurs in the deep infrared, the scaling behavior comes with distinctive properties such as critical slowing down and universality characteristic for critical phenomena. Our analysis shows that this far-from-equilibrium critical dynamics can be understood as a strong turbulence phenomenon. 

Based on our results we propose to study turbulence phenomena in dilute ultracold gases with hindsight to the scaling properties of correlation functions, in particular of the momentum distribution of particles in the gas. It would be striking to find experimentally the predicted strong turbulence regime, which would have important consequences also in other areas of physics such as early-universe cosmology where similar phenomena can crucially determine the thermal history of our universe.

\acknowledgments \noindent 
J.B.\ would like to thank G. Hoffmeister, A.~Rothkopf, J.~Schmidt and D.~Sexty for collaboration on related work. T.G. would like to thank N.~Berloff, B.~Nowak, J.~M.~Pawlowski, and A.~Polkovnikov for inspiring and useful discussions. 
The authors would also like to thank KITP and the University of California at Santa Barbara for their hospitality, where part of this work was initiated. 
This research was supported in part by the National Science Foundation under Grant No. PHY05-51164.
T.G.\ would like to thank M.~Holland, JILA, and the University of Colorado at Boulder for their hospitality, and acknowledges support by the Deutsche Forschungsgemeinschaft, as well as by the Alliance Program of the Helmholtz Association (HA216/EMMI).

\begin{appendix}

\section{Notation}
\label{app:notation}
We chose the ($+---$) convention for the Minkowski metric. 
The Minkowski product between 4-vectors
$p = (p^0,p^1,p^2,p^3) = (p_0,\vec p) = (\omega,\vec p)$
and
$x = (x^0,x^1,x^2,x^3) = (x_0,\vec x) = (t,\vec x)$
is then
$  px = p_0x_0 - \vec p\cdot\vec x$.
Functions of 4-vectors are Fourier-transformed with the Minkowski product as follows:
$\mathcal{F}[f(p)](x) = f(x) = (2\pi)^{-4}\int{\mathrm{d}^4p}\exp\{-ipx\}f(p)$,
$\mathcal{F}[f(x)](p) = f(p) = \int\mathrm{d}^4x \exp\{ipx\}f(x)$.
In $d<3$ spatial dimensions all definitions are analogous.

Function letters denote both the function and its Fourier transform. 
The argument specifies which of the two is meant. 
The following convention is used for convolutions:
\begin{align}
  (f\ast g)(x) 
  &= \int\df{y}f(y)g(x-y),
\label{eq:notConvolution}
  \\
  (f\ast g)(p) 
  &= \int\frac{\df{q}}{(2\pi)^4}f(q)g(p-q).
\end{align}
The convolution theorem is then
\begin{align}
  \mathcal{F}[(f\ast g)](x) 
  &= (f\cdot g)(x) = f(x)\cdot g(x),
  \\
  \mathcal{F}[(f\ast g)](p) 
  &= (f\cdot g)(p) = f(p)\cdot g(p).
\end{align}
and the inverse versions are
$\mathcal{F}[(f\cdot g)](x)  =  (f\ast g)(x)$, 
$\mathcal{F}[(f\cdot g)](p)  =  (f\ast g)(p)$.
Arguments as often written as subscripts or dropped completely if clear from the context,
\begin{equation}
h(x) =  f(x)\cdot g(x) = f_x\cdot g_x = f\cdot g
\end{equation}
For compactness, matrix indices are sometimes written above function letters:
\begin{equation}
F_{ab}(p) = \stackrel{ab}{F}(p) = \stackrel{ab}{F}_p
\end{equation}
Contractions of the matrices $F$, $\rho$ etc. are written as follows:
\begin{equation}
F_{ab}\cdot\rho_{ab} = (F\cdot \rho).
\end{equation}
%

\section{The 2PI effective action approach to nonequilibrium dynamics}
\label{app:2PIEA-Approach}
In this appendix we review the aspects of the 2PI effective action approach to nonequilibrium dynamics.
For introductory texts see, e.g., Refs.~\cite{Berges:2004yj, Gasenzer2009a}.

\subsection{2PI effective action}
For a given initial-state density matrix $\rho_D(t_0)$ characterizing a system far from equilibrium, all information about the quantum field theory is contained in the generating functional for correlation functions:
\begin{align}
\label{eq:definingZneq}
  Z[J,K;\rho_D]
  &= {\rm Tr}\Big[\rho_D(t_0)\, {\cal T}_{\cal C}
  \exp\Big\{i \Big(\int_{x,{\cal C}}\! J_a^{\cal C}(x) \Phi_a(x)
  \nonumber\\
  & +\ \frac{1}{2} \int_{x y,{\cal C}}\!
  \Phi_a(x)R_{ab}^{\cal C}(x,y)\Phi_b(y)\Big)\Big\}\Big]  \, ,
\end{align}
with Heisenberg field operators $\Phi_a(x)$ obeying the non-relativistic commutation relations \eq{BoseCommutator}.
In Eq.~(\ref{eq:definingZneq}), ${\cal T}_{\cal C}$ denotes time-ordering along the closed time path $\cal C$ leading from
the initial time $t_0$ along the real time axis to some arbitrary time $t$ and back to $t_0$, with $\int_{x,{\cal C}} \equiv\int_{\cal C} {\mathrm d} x_0 \int {\mathrm d}^d x$. 
Contour time ordering along this path corresponds to usual time ordering along the forward piece ${\cal C}^+$ and antitemporal ordering on the backward piece ${\cal C}^-$. 
Note that any time on ${\cal C}^-$ is considered later than any time on ${\cal C}^+$. 
The source terms in Eq.~(\ref{eq:definingZneq}) allow to generate correlation functions by functional differentiation such as
\begin{align}
  \langle {\cal T_C}\Phi(x_1)\cdots\Phi(x_n)\rangle
  &= \left.\frac{\delta^nZ[J,K;\rho_D]}
                {i^n\delta J(x_1)\cdots\delta J(x_n)}
     \right|_{J,K\equiv0},
\label{eq:ClCorrFfromZ}
\end{align}
where the field indices have been suppressed, $x_{i}=(x_{0,i},\mathbf{x}_{i})$ is a $d+1$-dimensional coordinate, and we have used that for the closed time path $Z = 1$ in the absence of sources. 

In order to obtain the dynamical evolution of the system, i.e., the time dependence of the correlation functions \eq{ClCorrFfromZ} requires precise knowledge of the generating functional $Z$.
For most nontrivial practical applications it is, however, not feasible to directly compute $Z$, e.g., by use of Monte Carlo techniques.
In contrast to many cases of imaginary-time evolution in equilibrium this is prevented by a variant of the sign problem.
A possible way out is the reformulation of the problem in terms of an effective action functional.
The final aim of this approach is the derivation of time evolution equations for certain correlation functions of interest, in particular for the lowest order connected correlation functions or cumulants as defined in Eqs.~\eq{phi} and \eq{G}.

The functional derivative relations  \eq{ClCorrFfromZ} which determine these two correlation functions are rewritten into stationarity requirements of an effective action functional $\Gamma[\phi,G]$ with respect to its arguments $\phi$ and $G$.
$\Gamma$ is obtained by a double Legendre transform of the Schwinger functional $W[J,R]=-i\ln Z[J,R]$ with respect to the source fields $J$ and $R$ introduced in the definition \eq{ClCorrFfromZ} of the generating functional $Z$.
The result is the 2PI effective action 
\begin{equation}
\label{eq:Gamma2PI}
  \Gamma[\phi,G]=\Gamma^\mathrm{(1loop)}[\phi,G]+\Gamma_2[\phi,G],
\end{equation}
where 
\begin{equation}
\label{eq:Gamma2PI1loop}
  \Gamma^\mathrm{(1loop)}[\phi,G]
  = S[\phi]+\frac{i}{2}\mathrm{Tr}\left(\ln G^{-1}+G_0^{-1}G\right)+\mathrm{const.}
\end{equation}
denotes the one-loop part obtained in a saddle-point approximation of the generating functional $Z$ involving the classical action $S$ and the classical inverse propagator
\begin{equation}
\label{eq:G0}
  iG^{-1}_{0,ab}(x,y)=i\frac{\delta^2 S[\phi]}{\delta\phi_{a}(x)\delta\phi_{b}(y)}.
\end{equation}
The trace in \Eq{Gamma2PI1loop} includes sums over all internal and space-time indices.
The functional $S$ depends on the classical field only and defines the underlying model to be considered.
A particular class of models considered in this article will be defined in \Sect{DynEqs} below.

The stationarity requirements for $\Gamma$ with respect to $\phi$ and $G$, 
\begin{align}
\label{eq:StatCondsphi}
    \frac{\delta\Gamma[\phi,G]}{\delta\phi_a(x)}
  &= -J_a(x)-\int_y R_{ab}(x,y)\phi_b(y),
  \\
\label{eq:StatCondsG}
  \frac{\delta\Gamma[\phi,G]}{\delta G_{ab}(x,y)}
  &= -\frac{1}{2}R_{ab}(x,y),
\end{align}
are equivalent to the equations of motion of $\phi_{a}$ and $G_{ab}$.

If the initial state $\rho_{D}(t_{0})$ is taken to be Gaussian, then only the above introduced cumulants $\phi_{a}(t_{0},\mathbf{x})$ and $G_{ab}(t_{0},\mathbf{x};t_{0},\mathbf{y})$ are nonzero at $t=t_{0}$. 
Then, the initial state $\rho_{D}(t_{0})$, in the generating functional $Z$, can be absorbed into the sources $J$ and $R$ at $t=t_{0}$.
As a consequence, these sources can be set identically to zero in the equations of motion \eq{StatCondsphi},  \eq{StatCondsG} which thus close and describe the evolution of an isolated system starting in a state defined only by the first and second order cumulants at initial time $t_{0}$.

\subsection{Dynamic equations}
\label{app:DynEqs}
The effective-action approach defined in Eqs.~\eq{Gamma2PI}, \eq{Gamma2PI1loop} is, so far, valid for a general class of models describing the evolution of a nonrelativistic bosonic field $\Phi_{a}(x)$.
In Eqs.~\eq{Sclassphi4}, \eq{G0inv} we have specified the particular model by defining the classical action functional $S$.

Given the classical action, the contribution $\Gamma_{2}[\phi,G]$ to the 2PI effective action, see \Eq{Gamma2PI}, is defined in terms of an infinite series of all possible 2PI diagrams without external legs, formed from the bare 4-vertex defined by the interaction part in \Eq{Sclassphi4}, full propagators $G_{ab}(x,y)$, and classical field insertions $\phi_{a}(x)$, i.e., of all such diagrams which do not fall apart on opening at most two lines $G$.
From the resulting 2PI effective action the dynamic equations are determined by functional differentiation as defined in Eqs.~\eq{StatCondsphi} and \eq{StatCondsG}.
Since in general $\Gamma_{2}$ is an infinite series, also the resulting equations contain an infinite number of terms of increasing order in the number of bare couplings, internal lines $G$ and fields $\phi$.
To obtain a practically solvable set of equations the diagrammatic series $\Gamma_{2}$ needs to be truncated as discussed in more detail in \Sect{NLO1N} in the main text.

While a more detailed account of the standard procedures summarized above is beyond the scope of this article we point out that the principal reason for deriving the equations of motion for $\phi$ and $G$ by use of a stationarity condition is that the resulting equations conserve particle number and energy irrespective of the particular truncation of $\Gamma_{2}$.
In any such truncation the equations of motion imply a totally conserved particle current density $j_{\kappa}(x)=(\rho(x),\mathbf{j}(x))$ as well as energy momentum tensor $T_{\kappa\lambda}(x)$ with $\partial^{\kappa}j_{\kappa}(x)=0$ and $\partial^{\kappa}T_{\kappa\lambda}(x)=0$, respectively.
Integrated over space, these relations describe the conservation of total particle number and energy.
This important feature forms a necessary requirement for the resulting equations to be applicable to long-time dynamics, i.e., most notably, not to lead to secular evolution in this limit.

\section{2PI self energy}
\label{app:Sigma}

From the 2PI effective action the self energies \eq{Sigma0Frho} in two-loop order, for $\phi\equiv0$, are found as
\begin{align}
  &\left(\begin{array}{r}
        \Sigma^F_{ab}(x,y) \\ -\frac{1}{2}\Sigma^\rho_{ab}(x,y)
	\end{array}\right)
   = -\frac{2g}{\cal N}\Bigg[\left(\begin{array}{r}
        R_{ab}^F(x,y) \\ -\frac{1}{2}R_{ab}^\rho(x,y)
	\end{array}\right) 
  \nonumber
\end{align}
\begin{align}
  &\quad 
  +\ \left(\begin{array}{rr}
        \Pi^F(x,y) & 
	\frac{1}{2}\Pi^\rho(x,y) \\
       -\frac{1}{2}\Pi^\rho(x,y) & 
        \Pi^F(x,y) \end{array}\right)
    \left(\begin{array}{r}
        F_{ab}(x,y) \\ -\frac{1}{2}\rho_{ab}(x,y)
	\end{array}\right)\Bigg],
\label{eq:Sigma2loop}
\end{align}
with ($F^2=F_{ab}F_{ab}$, etc.)
\begin{align}
  &\left(\begin{array}{r}
        \Pi^F(x,y) \\ \Pi^\rho(x,y)
	\end{array}\right)
  =
  \frac{g}{\cal N}
  \left(\begin{array}{c}
        F(x,y)^2-\frac{1}{4}\rho(x,y)^2 \\
	2F_{ab}(x,y)\rho_{ab}(x,y)
	\end{array}\right),
\label{eq:PiFrho}
\end{align}
and
\begin{align}
  &\left(\begin{array}{r}
        R^F \\ R^\rho
	\end{array}\right)
  =
  \frac{2g}{\cal N}
  \left(\begin{array}{c}
        F^3-\frac{1}{4}(F\rho^2+\rho F\rho+\rho^{2} F) \\
	F^{2}\rho+F\rho F+\rho F^{2} -\frac{1}{4}\rho^3
	\end{array}\right),
\label{eq:RFrho}
\end{align}
where $(F^{3})_{ab}(x,y)=F_{ac}(x,y)F_{dc}(x,y)F_{db}(x,y)$, $(F\rho^{2})_{ab}$ $(x,y)=F_{ac}(x,y)\rho_{dc}(x,y)F_{db}(x,y)$, etc.

For the 2PI effective action in NLO of the expansion in $1/\mathcal{N}$, $\Gamma_2[\phi,G]=\Gamma_2^\mathrm{LO}[\phi,G]+\Gamma_2^\mathrm{NLO}[\phi,G]$ the diagrammatic expansion of which is shown in \Fig{DiagrExpGamma2NLO} we obtain, using \Eq{SigmafromGamma2}, the self energies \eq{Sigma0Frho} as
\begin{align}
  &\left(\begin{array}{r}
        \Sigma^F_{ab}(x,y) \\ -\frac{1}{2}\Sigma^\rho_{ab}(x,y)
	\end{array}\right)
   = -\frac{2g}{\cal N}\Bigg[\left(\begin{array}{r}
        I^F(x,y) \\ -\frac{1}{2}I^\rho(x,y)
	\end{array}\right)
        \phi_a(x)\phi_b(y) 
  \nonumber\\
  &\quad 
  +\ \left(\begin{array}{rr}
        \Delta^F(x,y) & 
	\frac{1}{2}\Delta^\rho(x,y) \\
       -\frac{1}{2}\Delta^\rho(x,y) & 
        \Delta^F(x,y) \end{array}\right)
    \left(\begin{array}{r}
        F_{ab}(x,y) \\ -\frac{1}{2}\rho_{ab}(x,y)
	\end{array}\right)\Bigg],
\label{eq:SigmaNLO1N}
\end{align}
where $\Delta^{F,\rho}(x,y)=I^{F,\rho}(x,y)+P^{F,\rho}(x,y;I^{F,\rho})$.
The resummation to NLO in $1/\cal N$ is taken care of by the coupled integral equations for $I^{F,\rho}$ \cite{Berges:2004yj}:
\begin{align}
  \left(\begin{array}{r}
        I^F(x,y) \\ I^\rho(x,y)
	\end{array}\right)
  &=
  \left(\begin{array}{c}
         \Pi^F(x,y) \\
         \Pi^{\rho}(x,y)
	\end{array}\right)
    \nonumber\\
   &\quad
   -\ \int_{t_0}^{x_0}\mathrm{d}z\,I^\rho(x,z)
      \left(\begin{array}{c}
         \Pi^F(z,y) \\
         \Pi^{\rho}(z,y)
	    \end{array}\right)
    \nonumber\\
   &\quad
   +\ \int_{t_0}^{y_0}\mathrm{d}z\,
      \left(\begin{array}{r}
            I^F(x,z) \\ 
            I^\rho(x,z) 
	    \end{array}\right)\Pi^{\rho}(z,y).
\label{eq:IFrho}
\end{align}
The functions $P^{F,\rho}$, which contribute to $\Delta^{F,\rho}$ in the self energies \eq{SigmaNLO1N} and vanish if $\phi_i\equiv0$, read \cite{Berges:2007ym}
\begin{align}
  &P^F(x,y;I^{F,\rho})
  = -\frac{2g}{{\cal N}}\Big\{
     H^F(x,y)
  \nonumber\\
  &\ 
  +\int_{t_0}^{y_0}\mathrm{d}z\left[H^F(x,z)I^\rho(z,y)+I^F(x,z)H^\rho(z,y)\right]
  \nonumber\\
  &\ 
  -\int_{t_0}^{x_0}\mathrm{d}z\left[H^\rho(x,z)I^F(z,y)+I^\rho(x,z)H^F(z,y)\right]
  \nonumber\\
  &\ 
  -\int_{t_0}^{x_0}\mathrm{d}v\int_{t_0}^{y_0}\mathrm{d}w\, I^\rho(x,v)H^F(v,w)I^\rho(w,y)
  \nonumber\\
  &\ 
  +\int_{t_0}^{x_0}\mathrm{d}v\int_{t_0}^{v_0}\mathrm{d}w\, I^\rho(x,v)H^\rho(v,w)I^F(w,y)
  \nonumber\\
  &\ 
  +\int_{t_0}^{y_0}\mathrm{d}v\int_{v_0}^{y_0}\mathrm{d}w\, I^F(x,v)H^\rho(v,w)I^\rho(w,y)\Big\},
\label{eq:PF}
\end{align}
\begin{align}
  &P^\rho(x,y;I^{F,\rho})
  = -\frac{2g}{{\cal N}}\Big\{
     H^\rho(x,y)
  \nonumber\\
  &\ 
  -\int_{y_0}^{x_0}\mathrm{d}z\left[H^\rho(x,z)I^\rho(z,y)+I^\rho(x,z)H^\rho(z,y)\right]
  \nonumber\\
  &\ 
  +\int_{y_0}^{x_0}\mathrm{d}v\int_{y_0}^{v_0}\mathrm{d}w\, I^\rho(x,v)H^\rho(v,w)I^\rho(w,y)\Big\},
\label{eq:Prho}
\end{align}
wherein the functions $H^{F,\rho}$ are defined as
\begin{align}
  H^F(x,y)
  &= -\phi_a(x)F_{ab}(x,y)\phi_b(y),
  \nonumber\\
  H^\rho(x,y)
  &= -\phi_a(x)\rho_{ab}(x,y)\phi_b(y).
\label{eq:HFHrho}
\end{align}
%

\section{Stationarity condition}
\subsection{General discussion}
\label{app:StatCond}
In this appendix we present a proof of \Eq{stationarity}: 
Consider the dynamic equation \eq{EOMF} for translationally functions $F$, $\rho$, and $\phi$, obeying \eq{invariant}, with $y$ set to zero:
\begin{align}
  &[ i\sigma^{2}_{ac}\partial_{x_0} + M_{ac} ] F_{cb}(x) 
  = -\int_{-\infty}^{x_0}\df{z}\Sigma^\rho_{ac}(x-z)F_{cb}(z) 
  \nonumber\\
  &\qquad
  +\ \int^0_{-\infty}\df{z}\Sigma^F_{ac}(x-z)\rho_{cb}(z)
  \label{eq:EOMFt0minusinfinity}
\end{align}
with
\begin{align}
  M_{ac} 
  &= \delta_{ac}\left[-\frac{\nabla^2_{\vec x}}{2m} + \frac{g}{2}(\phi_d \phi_d + F_{dd}(0))\right] 
  \nonumber\\
  &\quad
  +\ g(\phi_a \phi_c + F_{ac}(0)).
\end{align}
Contracting over indices $a=b$, and using the symmetry properties $\sigma^{2}_{ab} = -\sigma^{2}_{ba}$,  $F_{ab}(x) = F_{ba}(-x)$, $\dot{F}_{ab}(x) = -\dot{F}_{ba}(-x) $ and $(\partial_{j}^{2} F_{ab})(x) = (\partial_{j}^{2} F_{ba})(-x)$, one finds that the left-hand side of \Eq{EOMFt0minusinfinity} is invariant under $x\to-x$.
Considering the right-hand side of \Eq{EOMFt0minusinfinity}, and using the further symmetry properties $\rho_{ab}(-x)=-\rho_{ba}(x)$, $\Sigma^F_{ab}(-x) = \Sigma^F_{ba}(x)$ and $\Sigma^\rho_{ab}(-x) = -\Sigma^\rho_{ba}(x)$ following from Eqs.~\eq{SigmaNLO1N}-\eq{HFHrho} one finds, adding \Eq{EOMFt0minusinfinity} and minus its counterpart with the sign of $x$ reversed and summing over $a=b$, that
\begin{equation}
0 =  \int\df{z}\left(\Sigma^\rho_{ab}(x-z)F_{ba}(z)  -  \Sigma^F_{ab}(x-z)\rho_{ba}(z)\right)
\label{eq:RHSFxminusRHSFminusx}
\end{equation}
which, by the convolution theorem, proves the stationarity condition \eq{stationarity}.
Following the above line of argument for the dynamic equation \eq{EOMrho} for $\rho$ immediately shows that the condition corresponding to \Eq{RHSFxminusRHSFminusx} is automatically fulfilled.

\subsection{Field dependent contribution to $J$}
\label{app:PiFrho}
In this appendix we first sketch the derivation of Eqs.~\eq{PFDelta} and \eq{PrhoDelta} and finally quote the momentum-integral expressions for $J^{\lambda}$ \eq{JlambdaCompact} and $J^\Lambda$ \eq{JDeltaCompact}.
In the translationally invariant case, cf.~\Eq{invariant}, the integrals $P^{F}$ and $P^{\rho}$ can be written, using the notation introduced in \App{notation} and \Sect{vanishingfield}, as
\begin{align}
 P^{F}(x)
  =  \lambda\phi_a&\phi_b
         \big[ F_{ab} + F_{ab}\ast I^{A} + I^F\ast G^{A}_{ab}   
 \nonumber \\
 &\quad  -\ G^{R}_{ab}\ast I^F - I^{R}\ast F_{ab} 
 \nonumber \\
 &\quad  -\ I^{A}\ast F_{ab}\ast I^{R} 
 \nonumber \\
 &\quad  +\ I^F\ast G^{R}_{ab}\ast I^{R}  
 \nonumber \\
 &\quad  +\  I^{A}\ast G^{A}_{ab}\ast I^F 
 \big],
\label{eq:PFapp}
\\
 P^\rho(x) 
  =  \lambda\phi_a&\phi_b \big[ 
  \rho_{ab}- G^{R}_{ab}\ast I^\rho + \rho_{ab}\ast I^{A}  
 \nonumber \\
 &\quad  -\ I^{R}\ast\rho_{ab} + I^\rho\ast G^{A}_{ab}  
 \nonumber \\
 &\quad  +\ G^{R}_{ab}\ast I^\rho \ast I^{R}  
 \nonumber \\
 &\quad  -\ \rho_{ab}\ast I^{A}\ast I^{R} 
 \nonumber \\
 &\quad  +\ G^{A}_{ab}\ast I^{A}\ast I^\rho 
 \big],
\label{eq:PRapp}
\end{align}
with the retarded and advanced propagators
\begin{align}
  G^R_{ab}(x) 
   =  \theta\cdot\rho_{ab}, 
  &\qquad
  \label{eq:GR}
  G^A_{ab}(x) 
   =  \theta^-\cdot\rho_{ab}. 
\end{align}
The above expressions for $P^F$ and $P^\rho$ can be cast into a compact form with the help to $\lambda^\mathrm{eff}$ and $\Delta_{ab}$, to be defined below.
Consider the complex conjugate of \Eq{corelambda},
\begin{equation}
(1 + \theta^-\ast I^\rho)\cdot(1-\Pi^A) = 1
\end{equation}
obtained with $\Pi^R(p)^\ast = -\Pi^A(p)$, $I^\rho(p)^\ast = - I^\rho(p)$ and $\theta(p)^\ast = \theta^-(p)$.
Hence
\begin{align}
  \lambda^\mathrm{eff}(p) 
  & =  \frac{1}{(1+\Pi^R)(1-\Pi^A)} 
  = (1-\theta\ast I^\rho)(1+\theta^-\ast I^\rho). 
\end{align}
We furthermore define
\begin{align}
   \Delta_{ab}(p) 
   &=  2\Re\left[ \frac{G^R_{ab}}{1+\Pi^R} \right] = \frac{G^R_{ab}}{1+\Pi^R} + \frac{-G^A_{ba}}{1-\Pi^A}
\label{Delta} 
  \nonumber\\
  & =  (\theta\ast\rho_{ab}) - (\theta^-\ast\rho_{ba}) -\ (\theta\ast\rho_{ab})\cdot(\theta\ast I^\rho)  
  \nonumber\\
  &- (\theta^-\ast\rho_{ba})\cdot(\theta^-\ast I^\rho). 
\end{align}
Now, going over to momentum space and rearranging terms in Eqs.~\eq{PFapp} and \eq{PRapp}, one obtains Eqs.~\eq{PFDelta} and \eq{PrhoDelta}.

We close this appendix by quoting the momentum-integral expressions for $J^{\lambda}$, \Eq{JlambdaCompact}, and $J^\Lambda$, \Eq{JDeltaCompact}:
\begin{align}
  J^\lambda(p) 
   &= \frac{\lambda^2}{2(2\pi)^4}\phi_a\phi_b
   \int\df{k}\df{q}\delta_{p-k-q}
   \nonumber \\
  & \quad\times\ \Big[
   \lambda^\mathrm{eff}_p\cdot\big(
     \stackrel{ab}{\rho}_p\stackrel{cd}{F}_k\stackrel{cd}{F}_q 
   - \stackrel{ab}{F}_p\stackrel{cd}{\rho}_k\stackrel{cd}{F}_q 
   - \stackrel{ab}{F}_p\stackrel{cd}{F}_k\stackrel{cd}{\rho}_q   
   \big)
   \nonumber \\
  & \qquad 
  + \lambda^\mathrm{eff}_k\cdot\big(
    \stackrel{cd}{\rho}_p\stackrel{ab}{F}_k\stackrel{dc}{F}_q 
  - \stackrel{cd}{F}_p\stackrel{ab}{\rho}_k\stackrel{dc}{F}_q 
  - \stackrel{cd}{F}_p\stackrel{ab}{F}_k\stackrel{dc}{\rho}_q   \big)
  \nonumber \\
 & \qquad  
 + \lambda^\mathrm{eff}_q\cdot\big(
   \stackrel{cd}{\rho}_p\stackrel{dc}{F}_k\stackrel{ab}{F}_q 
 - \stackrel{cd}{F}_p\stackrel{dc}{\rho}_k\stackrel{ab}{F}_q 
 - \stackrel{cd}{F}_p\stackrel{dc}{F}_k\stackrel{ab}{\rho}_q   \big) 
 \Big],
\label{eq:Jlambda}
\\
  J^\Lambda(p)
  &= -\frac{\lambda^3}{2(2\pi)^8} \int\df{k}\df{q}\df{r}\delta_{p+k-q-r}\Lambda_{p+k}
   \nonumber \\
  & \quad\times\  \Big[ 
    \stackrel{ab}{\rho}_p\stackrel{ab}{F}_k\stackrel{cd}{F}_q\stackrel{cd}{F}_r 
  + \stackrel{ab}{F}_p\stackrel{ab}{\rho}_k\stackrel{cd}{F}_q\stackrel{cd}{F}_r
   \nonumber \\
  & \qquad \
  - \stackrel{ab}{F}_p\stackrel{ab}{F}_k\stackrel{cd}{\rho}_q\stackrel{cd}{F}_r 
  - \stackrel{ab}{F}_p\stackrel{ab}{F}_k\stackrel{cd}{F}_q\stackrel{cd}{\rho}_r \Big].
\label{eq:JDelta}
\end{align}
Here we have introduced a compact notation for matrix indices and momentum dependence, e.g.
\begin{equation}
  \stackrel{ab}{\rho}_p = \rho_{ab}(p).
\label{eq:rhoStackedIndices}
\end{equation}
Expression \eq{JDelta} can be combined with $J^{0}$, \Eq{J0}, to $J^{4}$, \Eq{J4} to
\begin{align}
  J^4(p)
  &= \frac{\lambda^2}{2(2\pi)^8} \int\df{k}\df{q}\df{r}\delta_{p+k-q-r}
  \nonumber \\
  &\times
  \lambda^{\mathrm{eff}}_{p+k}\cdot(1-\lambda\phi_{a}\phi_{b}\Delta_{ab})
   \nonumber \\
  & \quad\times\  \Big[ 
    \stackrel{ab}{\rho}_p\stackrel{ab}{F}_k\stackrel{cd}{F}_q\stackrel{cd}{F}_r 
  + \stackrel{ab}{F}_p\stackrel{ab}{\rho}_k\stackrel{cd}{F}_q\stackrel{cd}{F}_r
   \nonumber \\
  & \qquad \
  - \stackrel{ab}{F}_p\stackrel{ab}{F}_k\stackrel{cd}{\rho}_q\stackrel{cd}{F}_r 
  - \stackrel{ab}{F}_p\stackrel{ab}{F}_k\stackrel{cd}{F}_q\stackrel{cd}{\rho}_r \Big].
\label{eq:J4App}
\end{align}
%

\section{Scaling properties}
\label{app:ScalingProperties}
The scaling properties for $\rho$ and $F$ are given in Eqs.~\eq{scalingrho} and \eq{scalingF}, respectively.
In general, if two functions $f$ and $g$ scale like
\[
f(s^zp_0,s\vec p)=s^{-\gamma}f(p)\quad \textrm{and} \quad g(s^zp_0,s\vec p) = s^{-\delta}g(p),
\]
it follows that their convolution scales like
\begin{equation}
  (f\ast g)(s^zp_0,s\vec p) = s^{z+d-\gamma-\delta}(f\ast g)(p).
\label{eq:convolutionscalingNonrel}
\end{equation}
Hence, $\Pi^{\rho}(p) = \lambda(F\ast\rho)$, scales as
\begin{equation}
  \Pi^{\rho}(s^zp_0,s\vec p) = s^{z+d-4+2\eta-\kappa}\Pi^{\rho}(p),
\end{equation}

The Fourier-transformed step function $\theta(p)\sim\int\df{x} e^{ipx}\theta(x)$ scales as
\begin{equation}
  \theta(s^zp_0,s\vec p) = s^{-z-d}\theta(p).
\end{equation}
By the scaling rule for a convolution \eq{convolutionscalingNonrel}, we see that the retarded function $\Pi^R(p) = \theta\ast\Pi^{\rho}$ scales like $\Pi^{\rho}$:
\begin{equation}
  \Pi^R(s^zp_0,s\vec p) = s^{z+d-4+2\eta-\kappa}\Pi^R(p).
\label{eq:PiRscalingAppendix}
\end{equation}
If $\kappa>z+d-4+2\eta$, $|\Pi^R(p)| \gg 1$ in the IR, and one can neglect the 1 in the denominator of $\lambda^\mathrm{eff}=|1+\Pi^R|^{-2}$,
\begin{equation}
  \lambda^\mathrm{eff}(s^zp_0,s\vec p) = s^{2(\kappa+4-z-d-2\eta)}\lambda^\mathrm{eff}(p) \qquad \textrm{(IR)}.
\label{eq:scalinglambdaeffIR}
\end{equation}
In the UV limit, $|\Pi^R(p)| \ll 1$, thus
\begin{equation}
  \lambda^\mathrm{eff}(p) = 1 \qquad \textrm{(UV limit)}.
\label{eq:scalinglambdaeffUV}
\end{equation}

According to \Eq{convolutionscalingNonrel}, the retarded propagator \eq{GR}, $G^R_{ab}(p) =  \theta\ast\rho_{ab}$, scales like $\rho$:
\begin{equation}
  G^R_{ab}(s^zp_0,s\vec p) = s^{-2+\eta}G^R_{ab}(p),
\end{equation}
$\Delta_{ab}(p) = 2\Re[ G^R_{ab}/(1+\Pi^R)]$, \Eq{Delta}, as
\begin{align}
  \Delta_{ab}(s^zp_0,s\vec p) 
  &= s^{\kappa-d+2-z-\eta}\Delta_{ab}(p) \quad &\textrm{(IR)},
  \label{eq:scalingDeltaIR}
  \\
  \Delta_{ab}(s^zp_0,s\vec p) 
  &= s^{-2+\eta}\Delta_{ab}(p) \qquad &\textrm{(UV)}.
  \label{eq:scalingDeltaUV}
\end{align}
Finally, $\Lambda(p) = \varphi_a\varphi_b\Delta_{ab}\cdot\lambda^\mathrm{eff}$, \Eq{Lambda}, scales as
\begin{align}
  \label{eq:scalingLambdaIR}
  \Lambda(s^zp_0,s\vec p) 
  &= s^{3(\kappa - d-z)+5(2-\eta)}\Lambda(p)  \quad &\textrm{(IR)},\\
  \label{eq:scalingLambdaUV}
  \Lambda(s^zp_0,s\vec p) &= s^{-2+\eta}\Lambda(p)   \qquad &\textrm{(UV)}.
\end{align}
%

\section{Scaling transformations}
\label{app:Zakharov}
Scaling transformations are extensively used within the Kolmogorov-Zakharov theory of wave turbulence \cite{Zakharov1992a}. 
In this appendix we briefly describe the scaling transformation used in Sects.~\ref{sec:kappaUV} and \ref{sec:kappaIR} to derive scaling exponents.

The scaling transformation allows to exchange two variables in an integral expression despite the fact that one of them is a free variable, if the functions of the integrand obey a scaling property: 
Consider an integral which has the form of the scattering integral $J^{3}(p)$ integrated over the spatial momenta, $J^{3}(p_{0})=J^{\lambda}(p_{0})$, see Eqs.~\eq{Jlambda}, \eq{weakstationarityp0} ($p=(p_0,\vec p)$, etc.):
\begin{equation}
   I(p_0) = \int\limits_{k_0,q_0>0}\ddf{p} \df{k}\df{q} \delta(p-k-q) f(p) g(k) h(q),
\label{eq:Ip0}
\end{equation}
with $p_0>0$, and $f$, $g$ and $h$ obeying the scaling laws
\begin{align}
  f(s^zp_0,s\vec p)=s^{-\gamma}f(p_0,\vec p),\nonumber\\
  g(s^zp_0,s\vec p)=s^{-\delta}g(p_0,\vec p),\nonumber\\
  h(s^zp_0,s\vec p)=s^{-\epsilon}h(p_0,\vec p).
  \label{eq:Scalingfgh}
\end{align}
The following transformations allows to swap $p$ and $k$~\cite{Berges:2008wm,Berges:2008sr}:
\begin{align}
  p_0&=\frac{p_0}{k_0'}k_0'; 
  &k_0&=\frac{p_0}{k_0'}p_0; 
  &q_0&=\frac{p_0}{k_0'}q_0'
  \nonumber\\
  \vec p &= \left(\frac{p_0}{k_0'}\right)^\frac{1}{z}\vec k'; 
  &\vec k &= \left(\frac{p_0}{k_0'}\right)^\frac{1}{z}\vec p'; 
  &\vec q &= \left(\frac{p_0}{k_0'}\right)^\frac{1}{z}\vec q'
  \label{eq:Zakharovp0}
\end{align}
leading to the result
\begin{align}
  I(p_0) =& \int\limits_{k_0,q_0>0}\ddf{p}\df{k}\df{q} \delta(k-p-q) f(k) g(p) h(q) 
  \nonumber\\
  &\qquad\qquad\times\ 
  \left({p_0}/{k_0}\right)^{-\beta}
\end{align}
where
\begin{equation}
  -\beta = \frac{1}{z}(2d+2z-\gamma-\delta-\epsilon).
\end{equation}
Hence, the arguments $p$ and $k$ are exchanged at the cost of a factor $(p_{0}/k_{0}')^{-\beta}$ which involves the scaling exponents.
The scaling transformations applied in Sects.~\ref{sec:kappaUV} and \ref{sec:kappaIR} on the integrals contributing to $J^{4}(p_{0})$ which involve one more frequency-momentum integration are performed analogously.

\end{appendix}

\bibliography{bibtex/additions,bibtex/mybib}

\end{document}